\documentclass[10pt,journal,compsoc]{IEEEtran}
\ifCLASSOPTIONcompsoc
  \usepackage[nocompress]{cite}
\else
  \usepackage{cite}
\fi
\ifCLASSINFOpdf
\else
\fi

\usepackage{lineno,hyperref}
\usepackage{amsmath,amssymb,amsfonts}
\usepackage{algorithmic}
\usepackage{graphicx,float}
\usepackage{textcomp}
\usepackage{xcolor}
\usepackage{subcaption}
\usepackage{url}
\usepackage{mwe}

\usepackage{xspace}
\usepackage{adjustbox}
\usepackage{multirow}
\usepackage{array}
\usepackage{relsize}
\usepackage{booktabs}

\newif\ifdraft
\drafttrue

\begin{document}
%
\title{Watch out for Extrinsic Bugs! A Case Study of their Impact in Just-In-Time Bug Prediction Models on the OpenStack project}
%
%
%
%

\author{Gema~Rodr\'iguez-P\'erez,~\IEEEmembership{Member,~IEEE,}  Meiyappan~Nagappan,~\IEEEmembership{Member,~IEEE,} 
and~Gregorio~Robles,~\IEEEmembership{Senior Member,~IEEE}
	\IEEEcompsocitemizethanks{\IEEEcompsocthanksitem Gema Rodriguez-Perez and Meiyappan Nagappan are with the University of Waterloo, Waterloo, Canada.\protect\\
		E-mail: gema.rodriguez-perez, mei.nagappan\{@uwaterloo.ca\}
		\IEEEcompsocthanksitem Gregorio Robles is with the Universidad Rey Juan Carlos, Fuenlabrada, Madrid, Spain.\protect\\E-mail: grex@gsyc.urjc.es}}%

%
%

\markboth{IEEE Transactions on Software Engineering, May~2020}%
{Rodriguez-Perez \MakeLowercase{\textit{et al.}}: Watch out for Extrinsic Bugs!}
%



\IEEEtitleabstractindextext{%
\begin{abstract}

Intrinsic bugs are bugs for which a bug-introducing change can be identified in the version control system of a software. In contrast, extrinsic bugs are caused by external changes to a software, such as errors in external APIs; thereby they do not have an explicit bug-introducing change in the version control system. Although most previous research literature has assumed that all bugs are of \emph{intrinsic} nature, in a previous study, we show that not all bugs are intrinsic. This paper shows an example of how considering extrinsic bugs can affect software engineering research. Specifically, we study the impact of extrinsic bugs in Just-In-Time bug prediction by partially replicating a recent study by McIntosh and Kamei on JIT models. These models are trained using properties of earlier bug-introducing changes. Since extrinsic bugs do not have bug-introducing changes in the version control system, we manually curate McIntosh and Kamei’s dataset to distinguish between intrinsic and extrinsic bugs. Then, we address their original research questions, this time removing extrinsic bugs, to study whether bug-introducing changes are a moving target in Just-In-Time bug prediction. Finally, we study whether characteristics of intrinsic and extrinsic bugs are different. Our results show that intrinsic and extrinsic bugs are of different nature. 
When removing extrinsic bugs the performance is different up to 16 \% Area Under the Curve points. This indicates that our JIT models obtain a more accurate representation of the real world. We conclude that extrinsic bugs negatively impact Just-In-Time models. Furthermore, we offer evidence that extrinsic bugs should be further investigated, as they can significantly impact how software engineers understand bugs. 
\end{abstract}

}

\maketitle

\IEEEdisplaynontitleabstractindextext

%
\IEEEpeerreviewmaketitle

\IEEEraisesectionheading{\section{Introduction}\label{sec:introduction}}

\IEEEPARstart{R}ecent studies show that bugs do not have the same origin~\cite{rodriguez2018if,rodriguez2019how}. While some bugs have their origin in explicit changes recorded in the version control system (VCS) of the software, other bugs are due to external changes that are not recorded in the VCS, e.g., changes in external APIs, compatibility changes or even changes in the specifications.

Rodr\'iguez-P\'erez~\emph{et al.} distinguish between \emph{intrinsic bugs} and \emph{extrinsic bugs}. \emph{Intrinsic bugs} are those bugs that have an explicit \emph{bug-introducing change} (BIC) in the VCS. On the other hand, \emph{extrinsic bugs} do not have a BIC recorded in the VCS because there is no explicit change in the VCS of the project that introduced the bug~\cite{rodriguez2019how}. 
This may be because the bug was caused (i) by a change in the environment where the software is used, (ii) because requirements changed, (iii) in an external library used by the project, or (iv) by an external change to the VCS of the project, among other reasons.

In the case of extrinsic bugs, it is not possible to identify a BIC in the VCS for a given bug; therefore we cannot link the \emph{bug-fixing change} (BFC) to a BIC.
This finding can put in jeopardy the results of previous studies as software engineering research has always considered all bugs to be intrinsic. For instance, \emph{Just-In-Time} (JIT) bug prediction models~\cite{kamei2013large}, which are built at change-level, can be affected as they are built with the assumption that for each bug there is always a BFC and a BIC in the VCS~\cite{sliwerski2005changes,kim2006automatic, williams2008szz}.

Researchers have proposed different bug prediction techniques~\cite{giger2012method,hata2012bug,zimmermann2007predicting,nagappan2005use,hall2012systematic}. But, 
JIT bug prediction has many advantages over other bug predictions techniques~\cite{shihab2012industrial}. For example, JIT models allow developers to review risky changes at the time of being produced and they are built at a finer-granularity as changes are often smaller than modules.

To build JIT models and predict bugs before they are discovered in a software component, it is necessary to train these models using historical data of that software component and learn when a bug occurred in the past.
During the training phase, JIT models use datasets that connect bug reports with the code changes that fixed the bug (the BFC), and with previous code changes that introduced the bug in the software (the BIC). 

Then, to predict future bugs, JIT models use code change properties of BICs and BFCs, such as the size of the change, the number of files modified by the change, or the experience of the developer. Since extrinsic bugs cannot be linked to a BIC, we hypothesize that in JIT models the incorrect identification of BICs in extrinsic bugs may impact the quality of datasets used to train JIT models, and ultimately may impact JIT models themselves.

JIT models must be trained using reliable datasets to improve their performance and increase their trustworthiness~\cite{mockus2008missing,hall2012systematic}.
To do so, we need to identify extrinsic bugs and remove them from the dataset. Therefore, we can obtain reliable dataset that only contains intrinsic bugs, i.e., bugs for which we can identify a BIC.

To study the impact of extrinsic bugs in JIT models, we partially replicated a recent paper by McIntosh and Kamei that analyzed the performance of JIT models~\cite{mcintosh2018fix}.  Through this paper, we will refer to McIntosh and Kamei's paper as \emph{Mc\&K's paper} to improve readability. As in previous research, Mc\&K's paper considered all bugs to be intrinsic. 
To quantify the impact of extrinsic bugs on JIT models, we removed extrinsic bugs from their dataset.

Methodologically, to classify bugs as intrinsic or extrinsic, we followed the approach proposed in~\cite{rodriguez2018if, rodriguez2019how}, which requires manually analyzing the bugs and their textual information.
As this is a very labor-intensive task, we have focused only on one of the projects used as case study in Mc\&K's paper: OpenStack.
We first manually curated their dataset and classified 1,880 bugs as intrinsic or extrinsic. We then used this curated dataset to train JIT models removing extrinsic bugs and computed their performance when identifying future BICs. Finally, we compared Mc\&K's results with our results and deepened in the differences regarding intrinsic and extrinsic bugs. 


Thus, we analyze whether our manually curated dataset is different from Mc\&K's dataset (RQ1). We have therefore added a constraint (i.e., ``when extrinsic bugs are removed'') to Mc\&K original research questions to study the impact of extrinsic bugs in JIT models, (RQ2-RQ4). As we found a significant share of mislabeled bugs in Mc\&K's dataset, we also analyze what their impact is as well (RQ5). Mislabeled bugs refer to issue reports that have been considered as bug reports when, in fact, they are not reporting a bug but another software maintenance activities, e.g., enhancements or refactoring. Finally, we study if intrinsic, extrinsic, and mislabeled bugs have different characteristics (RQ6).

Our results indicate that (1) intrinsic and extrinsic bugs are different, (2) our manually curated dataset differs, in terms of number of bugs, over 40\% from an automatic extracted dataset, (3) JIT models obtain different performance in terms of Area Under the Curve (AUC) (up to 16\% AUC points) when they consider only intrinsic bugs, and (4) AUC scores are more stable after removing extrinsic bugs.


The remainder of this paper is organized as follows. 
Section~\ref{sec:rqs} presents the research questions. 
Section~\ref{sec:relatedwork} discusses related work.
Section~\ref{sec:study} describes the design of our case study, and Section~\ref{sec:model-construction} presents how the model is constructed and analyzed. 
Section~\ref{sec:results} presents the results.
Section~\ref{sec:discussion} discusses the findings, while Section~\ref{sec:threats} contains the threats to their validity. Finally, Section~\ref{sec:conclusions} draws conclusions.

\section{Research Questions}
\label{sec:rqs}

The research questions addressed in this work are:
\begin{itemize}
	\item \textbf{RQ1}: How does our manually curated dataset differ from the one by McIntosh and Kamei?
	
	\textbf{Motivation}: 
	JIT models should use as input intrinsic bugs, as BICs of extrinsic bugs cannot be identified in the VCS. Thus, we are interested in studying how different our manually curated dataset is compared to the dataset obtained automatically and used in McIntosh and Kamei.
	
	\textbf{Results}: Over 40\% of bugs could not be linked to a BIC: 11.3\% of the bugs in McIntosh and Kamei's dataset were classified as extrinsic bugs and 29.1\% as mislabeled issues.\\
	 
	\item \textbf{RQ2}: Do JIT models lose predictive power over time \emph{when extrinsic bugs are removed}?\\
	\textbf{Motivation}: McIntosh and Kamei found that JIT models that were trained with old source code properties of BICs lose predictive power. With this question we want to see how only considering intrinsic bugs affects the predictive power of the models.
	
	\textbf{Results}: 
	JIT models also lose predictive power after one year of being trained when only intrinsic bugs are considered.
	However, our JIT models obtained a different performance in terms of AUC values (up to 16\% AUC points) and a minor loss of predictive power for each period (up to 15\% AUC points). \\
	
	\item \textbf{RQ3}: How does the relationship between code change properties and the likelihood of BICs evolve in terms of time \emph{when extrinsic bugs are removed} ? \\
	\textbf{Motivation}: If code change properties\footnote{Code change properties are described in Table~\ref{taxonomy}.} of BICs change over time, the properties of prior and future BICs are different. McIntosh and Kamei studied this relationship and found that properties of BICs change through the evolution of project. However, as the dataset they used contained extrinsic bugs as well, we think that prior and future events may not have similar properties. Thereby the impact of code change properties might fluctuate.
	
	\textbf{Results}: We have found that the impact of code change properties is indeed different than the one reported in McIntosh and Kamei. When extrinsic bugs are removed, the code change properties related to the
	magnitude of the change (\emph{Size}) increase up to 18\% AUC points and the code changes properties related to the code review process (\emph{Review}) decrease up to 36\% AUC points.\\
	

	\item \textbf{RQ4}: How accurately do current importance scores of code change properties represent future ones \emph{when extrinsic bugs are removed}? \\
	\textbf{Motivation}: McIntosh and Kamei found that the importance scores for some of the most impactful code change properties are consistently under/overestimated.
	However, we think that the importance score of some properties might change over time when removing extrinsic bugs.
	
	\textbf{Results}: We found that the importance scores for some of the most impactful code change properties are consistently under/overestimated as well. However, the stability of property importance score remains similar in both short and long JIT period models -- in McIntosh and Kamei this only applies to short-term models.\\


	\item \textbf{RQ5}: How do mislabeled bugs affect JIT models?
	
	\textbf{Motivation}: While manually curating the dataset in McIntosh and Kamei, we found a considerable share of mislabeled bugs. In RQ2-RQ4, we considered them as part of the input data for the JIT models. With this RQ we want to quantify how they affect the JIT models. As reported in recent research literature~\cite{tantithamthavorn2015impact}, we expect mislabeled bugs to have a low effect on the results.
	
	\textbf{Results}: Contrary to our expectations, we have found that mislabeled bugs also impact JIT models. When comparing the results of including mislabeled and intrinsic bugs in the dataset that fed JIT models with the best results that these models can obtain, we saw that they lose up to 4\% AUC points. 
	\\
	
	\item \textbf{RQ6}: Are the properties of BFCs and BICs linked to extrinsic, intrinsic, and mislabeled bugs different? \\
	\textbf{Motivation}: The results from RQ2-RQ5 show an improvement in terms of AUC in JIT models when removing extrinsic bugs. To ensure that these results can be considered statistically significant, we need to analyze how different code change properties of BFCs and BICs linked to extrinsic and intrinsic bugs are.
	
	\textbf{Results}: Intrinsic and Extrinsic bugs present statistically significant differences in the code change properties of BFCs and BICs linked to them. Furthermore, these properties are more similar between extrinsic and mislabeled than between intrinsic and mislabeled bugs. 
	 

\end{itemize}

\section{Related Work}
\label{sec:relatedwork}

In this section, we contextualize our work with past studies on bug origins, JIT bug prediction models and mislabeling issues.

\subsection{Origin of Bugs}

JIT bug prediction models need to identify \emph{bug-fixing changes} (BFCs) and \emph{bug-introducing changes} (BICs) from historical data, and then use the code change properties of those BFCs and BICs to train the JIT bug prediction models. That way JIT models can point out buggy changes before they are discovered in the software. 

Traditionally, JIT bug prediction models use the algorithm proposed by Sliwersky, Zimmermann, and Zeller \emph{(SZZ)}~\cite{sliwerski2005changes} to identify past BFCs and BICs. SZZ is a popular algorithm in bug prediction~\cite{rodriguez2018reproducibility}.
It assumes the last change that \emph{touched} the line(s) fixed in a BFC introduced the bug~\cite{zeller2011causes,sliwerski2005changes,kim2006automatic,williams2008szz}. In short, SZZ is an algorithm that links the VCS and the issue tracking (ITS) system of a project to identify the BFCs and their associated BICs.

Some authors have highlighted the limitations of linking BFCs with BICs, since the origin of some bugs might not be related to the lines modified in the BFC that fix the bug.
German~\emph{et al.} investigated bugs that manifest themselves in unchanged parts of the software and their impact across the whole system~\cite{german2009change}.
Chen~\emph{et al.} studied the impact of dormant bugs (i.e., bugs that were introduced in a version of the software system, but they were not found until much later) on bug localization~\cite{chen2014empirical}.
Prechelt and Pepper observed that BFCs may touch non-buggy lines, and even when they touched those lines, the actual BIC may have been made earlier~\cite{prechelt2014software}.
Ahluwalia~\emph{et al.} investigated the extend to which defect datasets ignore some defects because they have not been fixed~\cite{ahluwalia2019snoring}

Recently, Rodr\'iguez-P\'erez~\emph{et al.} have analyzed in detail the origin of bugs and found that some BICs cannot be identified in the VCS of a project because the change that caused the bug was not recorded in the VCS.
The authors identified two types of bugs: (1) intrinsic bugs, i.e., bugs caused by explicit changes recorded in the VCS, and (2) extrinsic bugs, i.e., bugs caused by external factors or changes to the software, as for instance changes in an external API, or changes in the requirements~\cite{rodriguez2018if,rodriguez2019how}. 

\subsection{Just In Time Bug Prediction Models}

JIT bug prediction models identify risky software changes instead of risky files or packages.
Kamei~\emph{et al.} proposed for the first time the JIT quality assurance technique that predicts defects at change-level~\cite{kamei2013large}. Recent studies have demonstrated that JIT models obtain sufficient prediction accuracy to be applied in practice~\cite{nayrolles2018clever,tan2015online}.

JIT bug prediction models assume that code change properties of past BICs are similar to code properties of future BICs. Therefore, we can use JIT models to learn from the past and predict the future. To achieve good prediction accuracy in JIT models, researchers rely on a variety of code changes properties to predict future BICs. These properties can be derived from the changes themselves~\cite{nagappan2006mining,mockus2008missing}, from VCSs and ITSs~\cite{graves2000predicting,kim2008classifying,kamei2013large}, or from code review systems~\cite{mcintosh2014impact,kononenko2015investigating}. 

These properties have been used in previous studies~\cite{kononenko2015investigating, kamei2013large,kim2008classifying} and can be grouped into six families of code change properties according to McIntosh and Kamei~\cite{mcintosh2018fix}:
(i) the \emph{Size} family measures the magnitude of the change, (ii) the \emph{Diffusion} family measures the dispersion of the changes across each modified file, (iii) the \emph{History} family measures the bug proneness of prior changes to the modified files, (iv) the \emph{Author Experience} family measures the experience of the author of the change, (v) the \emph{Reviewer Experience} family measures the experience the code reviewer(s) of the change, and (vi) the \emph{Review} family measures characteristics of change in the code review process.

We decided to study JIT bug prediction models because (1) they present many advantages over other bug prediction techniques~\cite{shihab2012industrial}, (2) they perform with high prediction accuracy~\cite{nayrolles2018clever}, and (3) they are a more practical alternative to traditional bug prediction techniques~\cite{kamei2016studying}. Nowadays, JIT bug prediction models are the best models yielding actionable results in the current state of the art.

\subsection{Mislabeling Issues}

As far as we know, there are two approaches to distinguish mislabeled bugs from real bugs. The first one is a manual analysis using the classification rules proposed by Herzig~\emph{et al.}~\cite{herzig2013s}. The second one is an automatic approach that uses regular expressions to identify (real) bugs from the commit messages of the BFCs. The SZZ algorithm implements this approach, so it has been widely used in previous research~\cite{mcintosh2018fix,da2017framework,kim2006automatic}.


Although the automatic approach can be quicker and easier than the manual
analysis, it may lead to noise in the dataset as some issue reports can be mislabeled, i.e., issue reports that describe defects but were not classified as such (or vice versa). Previous studies have shown the importance of correctly collecting data from VCS. Aranda and Venolia found that VCS and ITS hold incomplete or incorrect data~\cite{aranda2009secret}, which cause mislabeling data. Some studies have demonstrated that 33.8\%~\cite{herzig2013s} to 40\%~\cite{rodriguez2016bugtracking} of the bugs in the ITS are mislabeled.

This mislabeling might impact the performance of bug prediction models.
Kim~\emph{et al.} found that bug prediction models are considerably less accurate when they are trained using datasets that have a 20\%-35\% mislabeling rate~\cite{kim2011dealing}.
Herzig~\emph{et al.} observed that 33.8\% of all issue reports were mislabeled, and that this impacted the prioritization of files in bug prediction~\cite{herzig2013s}. Seiffert~\emph{et al.} carried out a comprenhensive study~\cite{seiffert2014empirical} that confirms Kim~\emph{et al.}'s findings~\cite{kim2011dealing}.

More recent studies suggest that mislabeled issues might not be a severe threat in bug prediction models since the mislabeling may not be necessarily random. For example, it is more likely that a novice developer mislabels an issue than an experience developer. Tantithamthavorn~\emph{et al.} found that precision is rarely impacted by mislabeled issues but recall is often impacted~\cite{tantithamthavorn2015impact}. They claim that the differences with Herzig~\emph{et al.}~\cite{herzig2013s} may be explained by the differences in their defect prediction experiments. Rahman~\emph{et al.} found that the number of buggy modules has a higher impact on bug model performance than the mislabeling data~\cite{rahman2013sample}. However, both studies agree that cleaning the data before training the models allows to achieve a better identification of indeed buggy modules.

Thus, to shed some lights on this topic, we decided to study how mislabeled bugs impact JIT bug prediction models. While previous studies looked at mislabeling in bug prediction models at the \emph{file or module} level, our study focuses on the \emph{change level}.

As far as we know, all previous studies about mislabeling have not differentiated between extrinsic and intrinsic bugs, considering them together. Therefore, there is no overlapping between what they considered as mislabeled bugs and what we refer to as extrinsic bugs in this work.

\section{Case study and Method}
\label{sec:study}

In this section, we describe our rationale for selecting the studied system and the data extraction process.

\subsection{Studied System: OpenStack}
\label{sec:OpenStack}

A qualitative analysis is required to ensure the correct identification of extrinsic bugs.
The output of this analysis is a manually curated dataset.
Creating this dataset is very labor intensive, since for every issue it is necessary to understand either the textual information in the issue report and the source code in the bug-fixing change, if not both.
Given this considerable effort, we selected one of the two case studies from Mc\&K's paper to partially replicate their study and understand the impact that extrinsic bugs have on JIT bug prediction models.

We chose OpenStack because
we are more familiar with OpenStack --in our previous study~\cite{rodriguez2018if} we investigate Nova, a component of OpenStack-- than with Qt.
Furthermore, we believe that OpenStack is an interesting and worthwhile project to study the impact of extrinsic bugs in JIT bug prediction models because it has more than 10,300 contributors with significant industrial support from several major IT companies such as Red Hat, Huawei, and IBM.
Currently, OpenStack has more than 330K commits with more than 48M lines of code and around 8,400 active developers.\footnote{http://stackalytics.com}
All its history is available and saved in a VCS (git), an ITS (Launchpad\footnote{https://launchpad.net/openstack}), and a source code review system (Gerrit\footnote{https://review.openstack.org/}).

\subsection{Data Extraction}
To study the impact of extrinsic bugs on JIT bug prediction models, we used the replication package\footnote{https://github.com/software-rebels/JITMovingTarget} provided by Mc\&K's paper~\cite{mcintosh2018fix}. With the information of the issues in the ITS and the VCS, we were able to manually annotated bugs on whether they were intrinsic or extrinsic.
We also found many issues that were wrongly considered bugs.

\begin{figure}[ht]
	\centering
	\includegraphics[width=\columnwidth, height=10em]{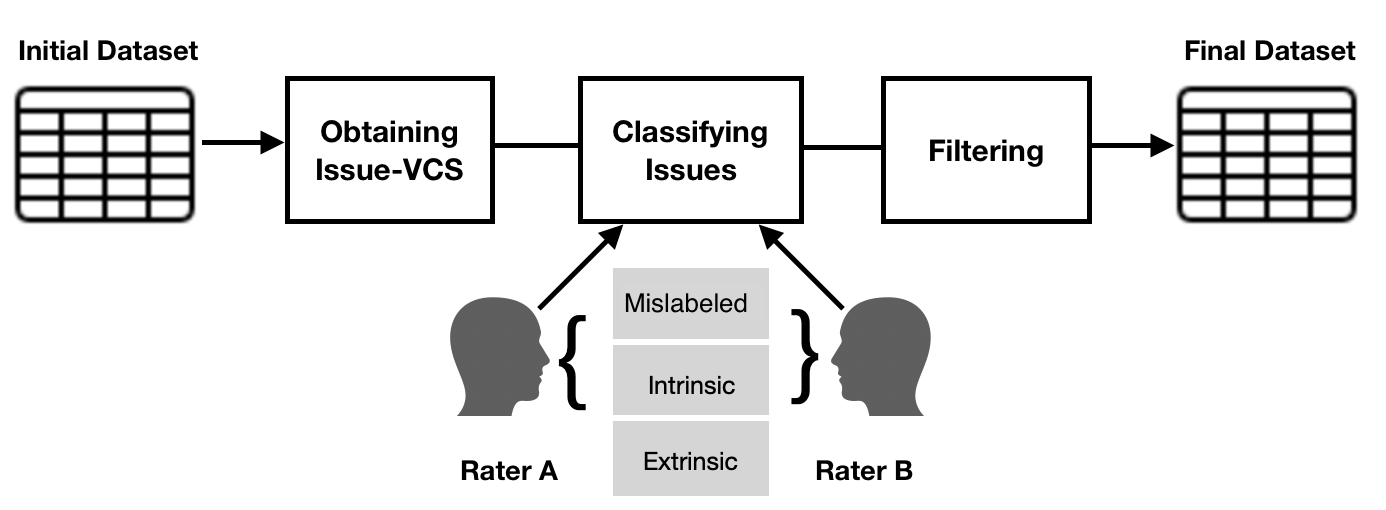}
	\caption{Overview of the steps followed to curate the dataset.}
	\label{fig:process}       
\end{figure}

Figure~\ref{fig:process} provides an overview of the phases followed to obtain our final dataset.
In the remainder of this section, we describe each phase in detail.

\subsubsection{Obtaining the Issue-VCS dataset}

The replication package only provides the final dataset to feed the JIT models studied in Mc\&K's paper.
We identified which of these changes were related to intrinsic or extrinsic bugs, and then removed the property of being a BIC when the change was an extrinsic bug.
For that, we required access to the issues in the ITS and their links to the changes in the VCS.
To ensure that we could address our research questions with the same dataset as Mc\&K, we asked them for the Issue-VCS dataset and obtained it.

The Issue-VCS dataset contains unique identifiers~\emph{(issueIDs)} of the issues and the timestamp when they were reported.
The issueIDs were used to link issues to code changes \emph{(changeIDs)}.
Thus, for each issueID there is one BFC and one or more changeIDs flagged as BICs.
In total, the Issue-VCS dataset contains 1,880 issueIDs linked to 1,904 changeIDs identified as BFCs, and 3,486 changeIDs identified as possible BICs.
Note that issueIDs and BFCs do not have to be related one-to-one, since an issueID can be fixed by more than one changeID.

\subsubsection{Classifying issues}

During the manual analysis, we noticed that the data used by Mc\&K not only contained extrinsic bugs, but issues that in fact were not bugs (e.g., other kind of issues such as request for new features or maintenance activities~\cite{herzig2013s}).

\begin{table}
	\renewcommand{\arraystretch}{1.3}
	\caption{First step: Classification rules for classifying issues as bug report or \emph{not a bug} report (\emph{mislabeled}).}
	\label{rules-bug-or-not}
	\centering
	
	\begin{smaller}
		\begin{tabular}{l}
			\toprule
			\hline
			An issue is classified as \textbf{not a bug} report (\emph{Mislabeled}) if ...
			\\
			(1) It reports a bug in test files.
			We assume that these bugs are caused by how \\ developers understand and test the code. 
			Thus, there is no change introducing \\ buggy code to source code of the project.
			 \\
			(2) It reports a clean up in the source code that does not interfere with \\ the performance 
			of the software.
			\\
			(3) It reports a misspelling or typo in the inline comments.\\
			(4) It reports a change in the source code to prevent future bugs.
			\\
			(5) The report has discordance in the comments between developers.
			\\
			(6) The report does not have a BFC.
			\\
			\hline
			An issue is classified as \textbf{bug report} if ...
			\\
			(1) It reports a misspelling or typo in the source code.
			\\
			(2) It reports that a previous change to the source code caused the bug.
			\\
			(3) It reports a buggy functionality implemented that should be known \\at the time of coding.\\
			(4) It reports an omission in the original code that should be considered \\at the time of coding.\\
			\hline
			\bottomrule 
		\end{tabular}
	\end{smaller}
\end{table}

Thus, first, following the guidelines provided by Herzig~\emph{et al.}~\cite{herzig2013s}, we manually classified the 1,880 issueIDs into \emph{bug report}, or \emph{not a bug} report (i.e., \emph{mislabeled}).
Table~\ref{rules-bug-or-not} shows the rules used in this first step.
Then, we manually classified the issues identified as bug reports as \emph{intrinsic} or \emph{extrinsic} bugs.
For doing so, we used the approach by Rodr\'iguez-P\'erez~\cite{rodriguez2018if}.
Table~\ref{rules-intrinsic-or-extrinsic} offers the specific rules used in this second step.

\begin{table}
	\renewcommand{\arraystretch}{1.3}
	\caption{Classification rules for classifying bug reports as intrinsic or extrinsic.}
	\label{rules-intrinsic-or-extrinsic}
	\centering
	
	\begin{smaller}
		\begin{tabular}{l}
			\toprule
			\hline
			A bug report is classified as \textbf{extrinsic} if ...
			\\
			(1) It reports a bug caused by a change in the environment \\ where the software is used.
			\\
			(2) It reports a bug because requirements have changed.
			\\
			(3) It reports a bug caused by an external change to the VCS of the project.
			\\
			(4) It reports a bug in an external library used by the project.
			\\
			\hline
			A bug report is classified as \textbf{intrinsic} if ...
			\\
			(1) There is no evidence to be classified as an extrinsic bug.\\
			\hline
			\bottomrule 
		\end{tabular}
	\end{smaller}
\end{table}

To remove subjectivity and bias in the classification, two raters having at least a master's degree in Computer Science manually classified the issueIDs.
The raters were individually trained in different stages, in each of them analyzing 100 random issueIDs from the data until they reached a near perfect agreement (0.81 - 1).
The ratio agreement between both raters was computed using Krippendorff's alpha, and the disagreements were resolved with online meetings.
After each stage, the raters discussed the discordance and added additional rationale to the guidelines.

Once the raters reached a near perfect agreement, they individually analyzed 25\% (470) of the issueIDs. At this point, the raters obtained a Krippendorff's alpha of 0.974 classifying issueIDs as a bug or not a bug, and a Krippendorff's alpha of 0.823 classifying bugs reports as extrinsic or intrinsic. We considered that the raters' agreement was high enough to analyze the remaining 1,410 issueIDs only by one rater (i.e., each rater classified 705 of the issueIDs).

The result of the classification procedure is a dataset where issues are labeled as (1) intrinsic bug, (2) extrinsic bug, or (3) not a bug (\emph{mislabeled}).


\subsubsection{Characteristics of the Changes}

Mc\&K extracted several code and review properties for each change from the VCS of OpenStack.
The properties were grouped in six \emph{families}: \emph{Size, Diffusion, History, Author Experience, Reviewer Experience, and Review}.
We use this information ``as is". The complete list of properties can be found in Table~\ref{taxonomy}.


\begin{table}
	\renewcommand{\arraystretch}{1.25}
	\caption{Taxonomy of changes provided by McIntosh and Kamei~\cite{mcintosh2018fix}.}
	\label{taxonomy}
	\centering
	\begin{smaller}
		\begin{tabular}{l|l|p{3.7cm}|c}
			\toprule
			\textbf{}&\textbf{Property} & \textbf{Description} & \textbf{Acron.} \\
			\hline
			\multirow{2}{4em}{Size}&Lines added & Number of lines added by the change. & la\\
			&Lines deleted & Number of lines deleted by the change. & ld\\
			\hline
			\multirow{2}{4em}{Diff.}&Subsystems & Number of modified subsystems. & ns\\
			&Directories & Number of modified directories. & nd\\
			&Files & Number of modified files. & nf\\
			&Entropy & Spread of modified lines across files. & ent\\
			\hline
			\multirow{2}{4em}{Hist.}&Unique Changes & Number of prior changes to the modified files. & nuc\\
			&Developers & Number of developers who have modified the file in the past. & ndev\\
			&Age & Time interval between the last and the current change.s & age\\
			\hline
			\multirow{2}{4 em}{ Author/\\ Reviewer\\ Exp.}&Prior Changes& The number of prior changes that an actor\footnote{Either the author or reviewer of a change} has participated\footnote{Either authored or reviewed} in. & aexp\\
			&Recent Changes & The number of \emph{aexp} weighted by the age of the changes. & arexp\\
			&Subs.Changes & Number of prior changes to the \emph{ns} that an actor has participated in.& asexp\\
			&Awareness & Proportion of \emph{aexp} to ns hat an actor has participated in.& asawr\\
			\hline
			\multirow{2}{4em}{Review}&Iterations & Number of times that a change was revised before integration. & nrev\\
			&Reviewers & Number of reviewers who have voted on integrating a change. & app\\
			&Comments & Number of non-automated, non-owner comments during the review of a change. & hcmt\\
			&Review Window & Time length between creation of a request and its final approval for integration.& rtime\\
			\hline
			\bottomrule 
		\end{tabular}
	\end{smaller}
\end{table}

\subsubsection{Final Dataset}
\label{finaldataset}

To obtain the final dataset that fed the JIT bug prediction models, Mc\&K merged the Issue-VCS dataset using changeID and issueID.
This merging filtered the dataset and mitigated false positives.
In addition, the dataset provided by Mc\&K (1) ignores potential BICs that only updated code comments or white spaces (an improvement to SZZ by Kim \emph{et al.}~\cite{kim2006automatic}); (2) filters out potential BICs that appear after the date that the implicated bug was reported~\cite{sliwerski2005changes}; and (3) ignores suspicious BFCs and suspicious BICs using the framework proposed by da~Costa~\emph{et al.}~\cite{da2017framework}.

Our goal is to study the impact of extrinsic bugs. Thus, we removed the changeIDs that were not BICs. Since extrinsic bugs do not have BICs, we removed the link between issueIDs classified as extrinsic bugs and their BICs following the recommendation of Rodr\'iguez-P\'erez~\emph{et al.}~\cite{rodriguez2019how}. Besides, we ignored changes that modified either at least 10,000 lines (``too much churn'') or 100 files (``too many files'') as they were likely no BICs.
The dataset obtained at this point is what we have called the \emph{final dataset}.
Finally, to study whether properties of BICs are consistent, we stratified the final dataset into periods of three and six months as Mc\&K's paper recommend~\cite{mcintosh2018fix}.
Table~\ref{filtering} shows the number of issues and BICs after each filtering phase.

\begin{table}
	\renewcommand{\arraystretch}{1.2}
	\caption{Number of unique issues and unique BICs that \emph{survive} each step of the filtering process.}
	\label{filtering}
	\centering
	\begin{smaller}
		\begin{tabular}{l|p{2.8cm} p{1cm} p{1cm}}
			\toprule
			\hline
			\textbf{\#} & \textbf{Filter} & \textbf{Issues} & \textbf{BICs} \\
			\hline
			$F_0$ & Issue-VCS dataset & 1,880 & 3,486  \\
			$F_1$ & Extrinsic Bugs & 1,668 &  2,925 \\
			$F_2$ & Too much Churn & 1,668& 2,920 \\
			$F_3$ & To many Files & 1,668&  2,911 \\
			$F_4$ & No lines added & 1,668&  2,907\\
			$F_5$ & Period & 1,668&  2,506 \\
			\hline
			\bottomrule 
		\end{tabular}
	\end{smaller}
\end{table}

Furthermore, in RQ5 (see~\ref{RQ5}) we discuss what the impact of removing mislabeled issues is. Thus, we added an additional filter to remove the links between issueIDs that were classified as mislabeled with their changeIDs identified as BFCs and BICs. 


\section{Model Construction and Analysis}
\label{sec:model-construction}

In this section, we describe the model construction and analysis approach.
Since we were partially replicating Mc\&K's paper, we exactly followed their model construction procedure. 
Thus, we used Mc\&K's design decisions with our final dataset, i.e., we did not modify any design decision from Mc\&K for the construction or analysis of the model.



\subsection{Model construction}


\subsubsection{Handling Collinear Properties}
Before constructing JIT models, we removed collinear code change properties to avoid distorting the modeled relationship between these code change properties and the likelihood of introducing bugs.

We used the Spearman rank correlation tests $\rho$ to remove code change properties that were highly correlated with one another.
For code change properties with correlation $|\rho|>0.7$, we only included one of the properties in the models.

Then, we fit preliminary models that explain each property using the others to remove redundant code change properties.
For that purpose, we used the \emph{redun} function available in the \emph{rms} R package.

\subsubsection{Fitting Regression Model}
Software Engineering researchers often use a nonlinear variant of multiple regression modeling to understand the relationship between software quality and software development practices~\cite{mcintosh2016empirical,zhou2011does}.
We fit JIT models using this technique as it relaxes the assumption of a linear relationship between the likelihood of introducing bugs and the code change properties; thus we can achieve a more accurate fit of the data.
We used restricted cubic splines, which fit smooth transitions at the points where curves change in direction, to fit our curves.

\subsection{Model Analysis}
\label{modelanalysis}

To answer our research questions, we analyzed the output of the JIT models using the different datasets.

\subsubsection{Analyzing the Performance of the Models (RQ2)}

The performance of JIT prediction models was assessed using two metrics: the Area Under Curve~(\emph{AUC}) and the Brier score.

The AUC is an evaluation metric for assessing the discriminatory power of a model, i.e., in our case its ability to differentiate between a BIC and not a BIC. 
AUC is calculated by measuring the area under the curve that plots the true positive rate of BICs against the false positive rate of BICs.
Its values range from 0 to 1; thus, the higher the AUC, the better the model is at predicting a BIC or not a BIC. 
When AUC is approximately 0.5, the model has no discrimination capacity, and it performs as random guessing.

The Brier score measures the calibration of the model.
It is computed by measuring the mean squared difference between the predicted probability assigned to the possible outcomes (being a BIC or not) for a change and its actual outcome.
The Brier score can range from 0 to 1; 0 indicates a perfect calibrated model, while 1 indicates the worst possible calibration for a model.

\subsubsection{Analyzing Property Importance (RQ3)}

Each of the six change properties families is comprised of several properties, and each property has been allocated with three degrees of freedom.
A model term represents each degree of freedom.
Thus, to estimate the impact that each family has on the explanatory power of the JIT models we jointly tested the set of model terms for each family using the Wald $\chi\textsuperscript{2}$ maximum likelihood tests~\cite{zhou2011does}.
We normalized the Wald $\chi\textsuperscript{2}$ values by the total Wald $\chi\textsuperscript{2}$ score of the JIT model to compare multiple models.
The larger the normalized Wald $\chi\textsuperscript{2}$ score, the more significant the impact a particular family of code change properties has on the explanatory power of our JIT models.

\subsubsection{Analyzing Property Stability (RQ4)}

To evaluate the stability of the importance scores for each family of code change properties $f$ over time, we calculated the difference between the importance scores of $f$ in a model that is trained using a period $n$ and a future model that is trained using a period $n + \chi$ where $\chi > 0$.


\section{Results}
\label{sec:results}

\subsection{ \textbf{RQ1: How does our manually curated dataset differ from the one by McIntosh and Kamei?}}


\textbf{Approach} Our manually curated dataset distinguishes among intrinsic bugs, extrinsic bugs, and mislabeled bugs, but Mc\&K's dataset does not. Thus, to further understand the differences between our manually curated dataset and an automatically extracted dataset (Mc\&K's paper), we computed the distributions and the probability density of the 1,880 issues in the two datasets. We used a kernel plot to present the distribution shape of the datasets. In kernel plots, wider sections represent a higher probability that members of the population will take on the given value; skinnier sections represent a lower probability.

\textbf{Results}
While manually curating the dataset, we identified 1,120 intrinsic bugs, 212 (11.3\%) extrinsic bugs, and 548 (29.1\%) mislabeled bugs.
We found that in Mc\&K's dataset, there are 1,413 BFCs linked to extrinsic bugs, 3,690 BFCs linked to intrinsic bugs, and 3,147 BFCs linked to mislabeled bugs. 

\begin{figure}[ht]
	\centering
	\includegraphics[width=0.48\textwidth]{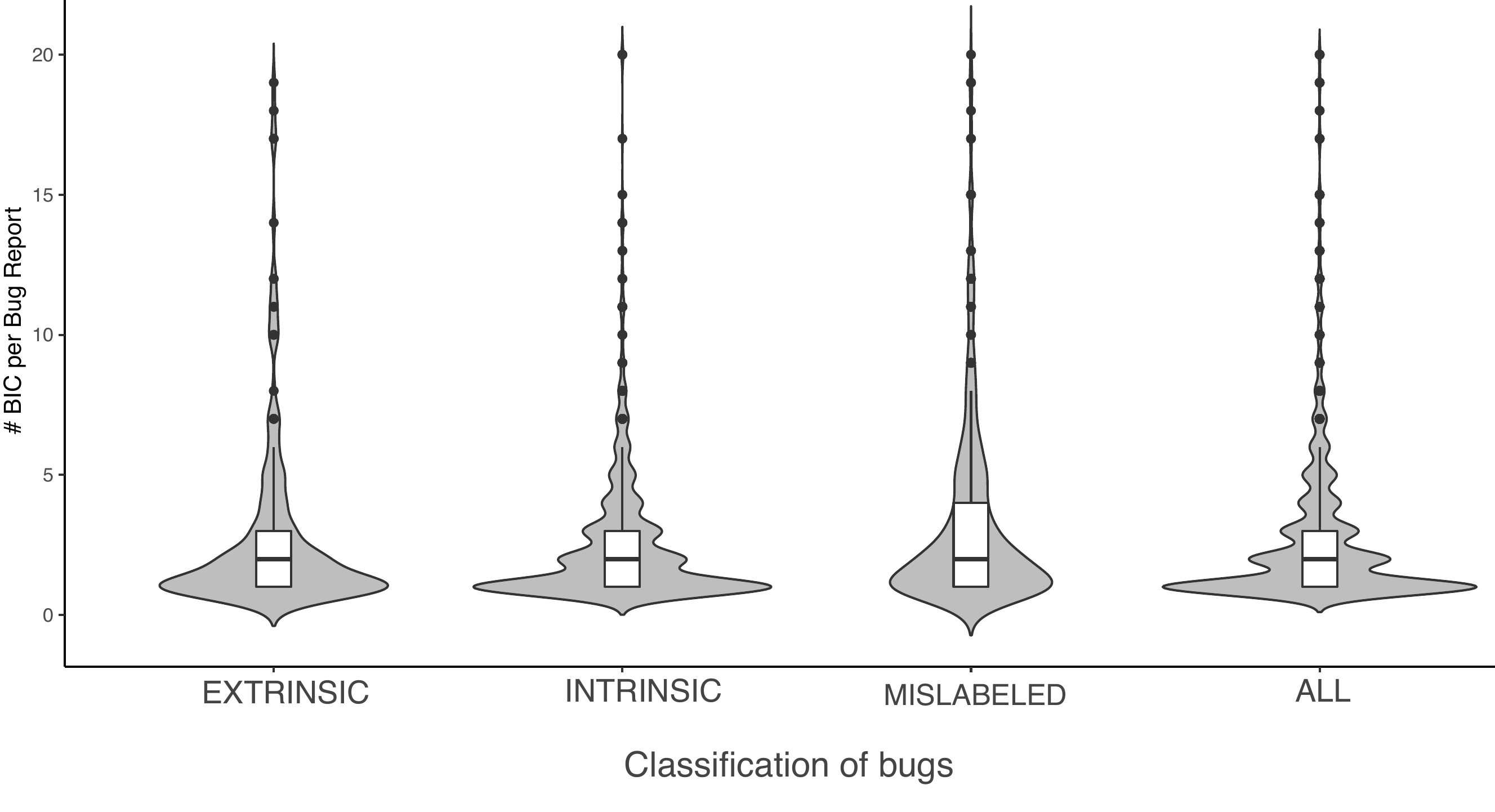}
	\caption{Distributions of the commits identified as BICs per bug report for each category of bug.}
	\label{fig:BIC}       
\end{figure}
 
Figure~\ref{fig:BIC} shows a violin plot with the distribution of the number of commits identified as BICs for each category. This figure offers evidence that (1) extrinsic, intrinsic, and mislabeled bugs have different distribution shapes; (2) Mc\&K's bugs (All) and intrinsic bugs have similar distribution shapes; and (3) the distribution shapes of extrinsic and mislabeled bugs differ from the one of intrinsic bugs. 


\vspace{0.2cm}
\fbox{\begin{minipage}{23em}
		\textbf{Answer to RQ1:}
		Over 40\% of the McIntosh and Kamei~\cite{mcintosh2018fix}'s dataset are not intrinsic bugs. Extrinsic and mislabeled bugs show different distribution shapes than intrinsic bugs.
\end{minipage}}


\subsection{\textbf{RQ2: Do JIT models lose predictive power over time when extrinsic bugs are removed?}}

To study how quick JIT models lose their predictive power we follow the same methodology as McIntosh and Kamei~\cite{mcintosh2018fix}. We split the data into periods, i.e., three-month and six-month periods of data. Then, we train the JIT models for each period and measure their performance on future periods.

\textbf{Approach}
Since older changes may have different characteristics than more recent ones, we used a short period model to train each period.
Short period models are JIT models trained only using changes that occurred during one time period, the latest one before the test period.
Since some studies suggest that the more training data, the better the results in bug detection models~\cite{rahman2013sample,zhang2016towards}, we also used long period models to train each period.
These long period models are JIT models trained using all the changes that occurred during or prior to the test period.

After training our JIT models in short and long periods, we measured their performance when they were applied in the test period.
The performance of our JIT models was measured using the AUC and the Brier score, as explained in Section~\ref{modelanalysis}.
For example, for training period 4, the short period model was trained using the changes in this period, and was tested using changes from period 5 onward; while the long period model was trained using changes in periods 1, 2, 3, and 4 and tested using period 5.
In both cases, the AUC and Brier measures were computed for each testing period individually.

\begin{figure}[tb]
	\centering
	\begin{subfigure}[b]{0.48\textwidth}
		\centering
		\includegraphics[width=\textwidth]{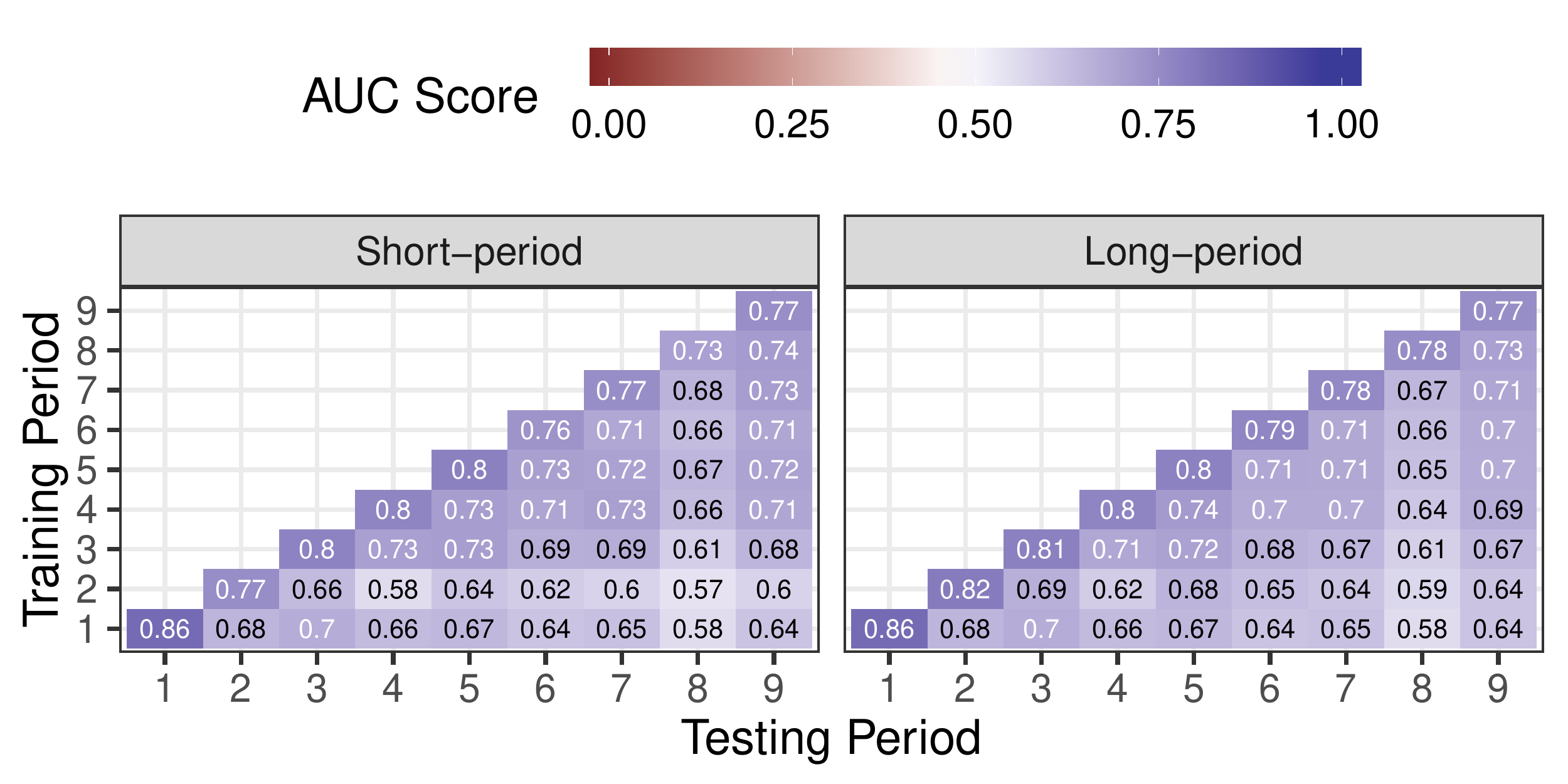}
		\caption[]%
		{{\small AUC in the three-month periods}}    
		\label{fig:AUCa}
	\end{subfigure}
	\hfill
	\begin{subfigure}[b]{0.48\textwidth}  
		\centering 
		\includegraphics[width=\textwidth]{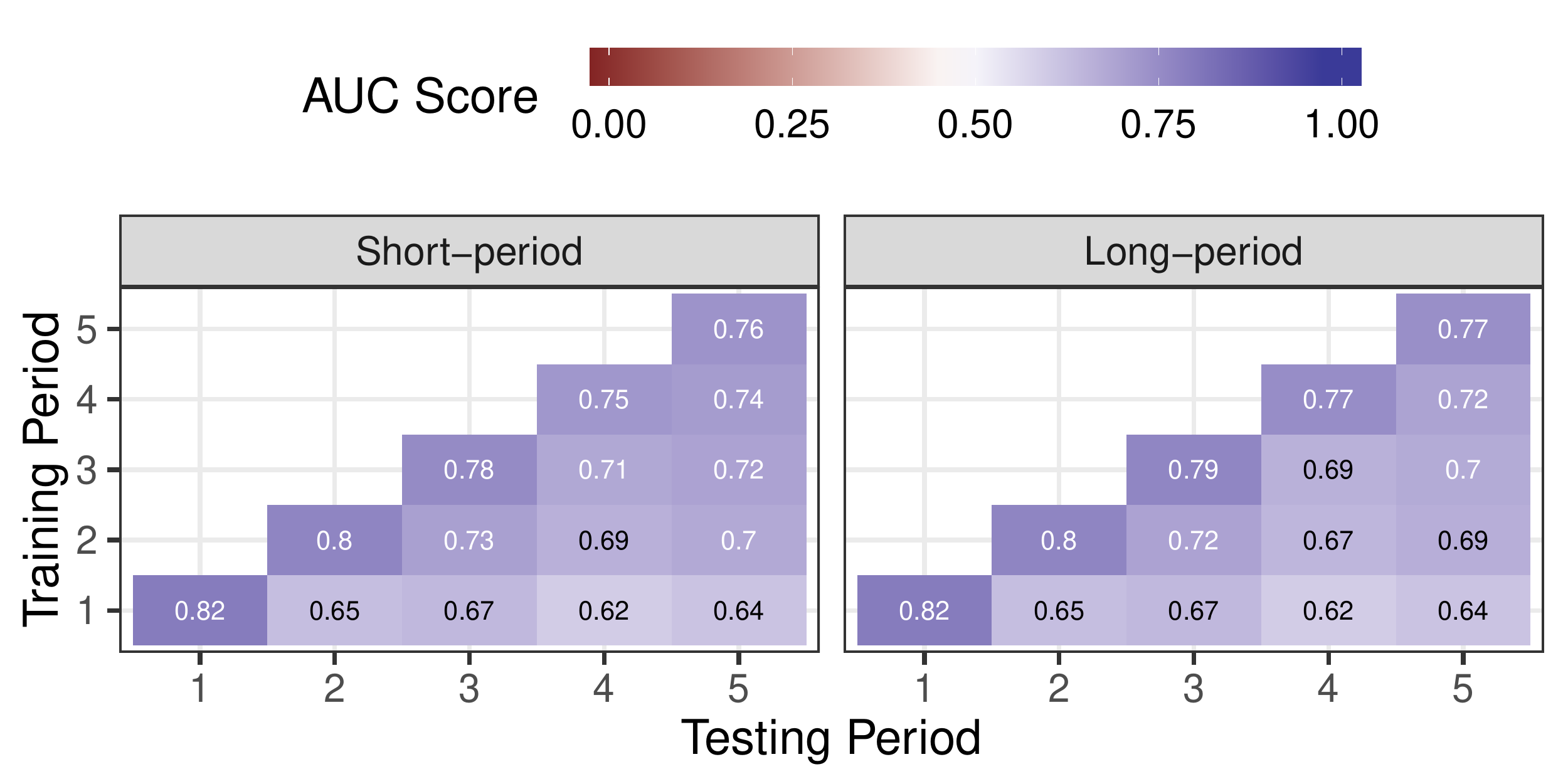}
		\caption[]%
		{{\small AUC in the six-month periods}}
		\label{fig:AUCb}
	\end{subfigure}
	\vskip\baselineskip
	\begin{subfigure}[b]{0.48\textwidth}
		\centering 
		\includegraphics[width=\textwidth]{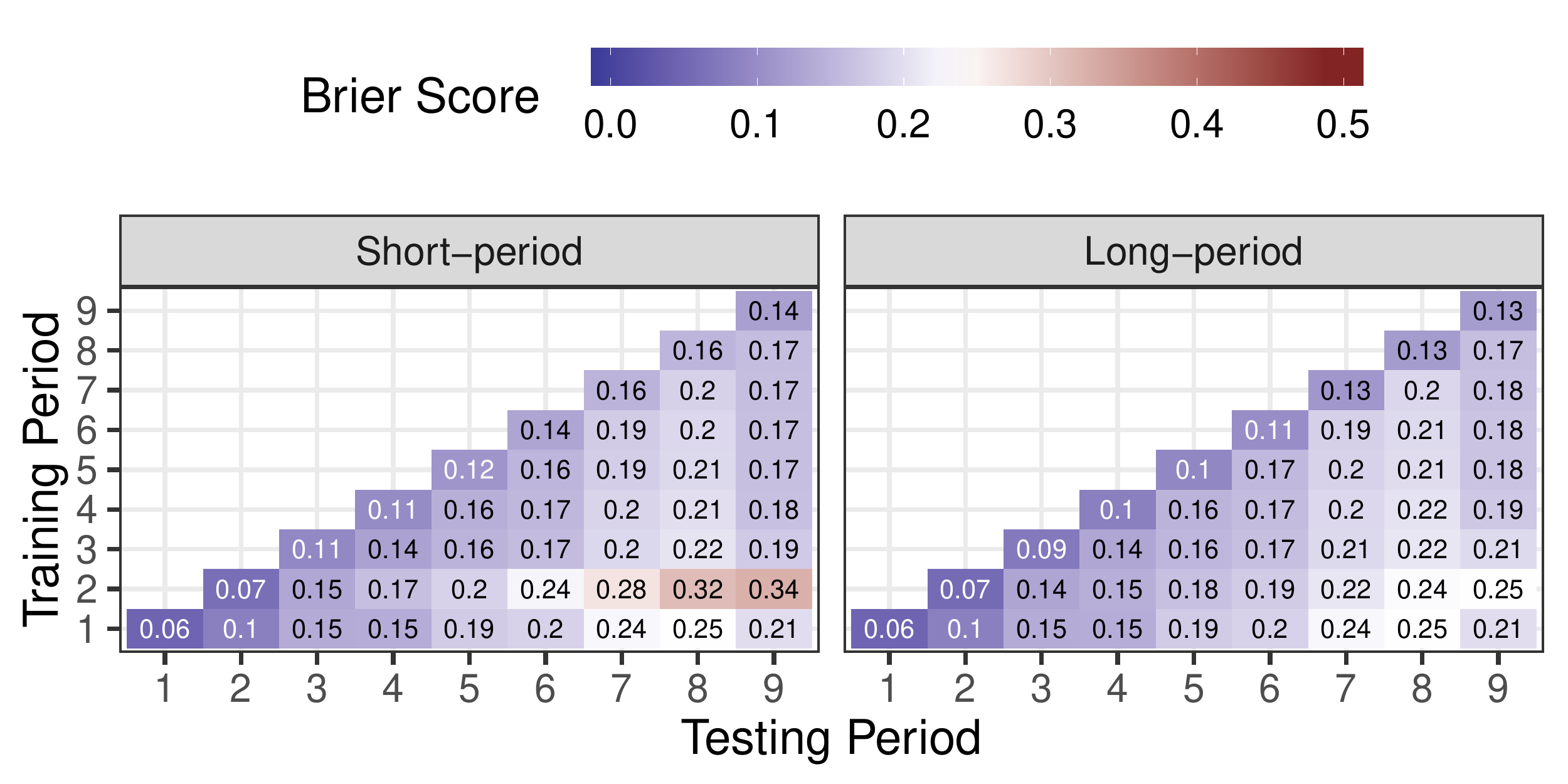}
		\caption[]%
		{{\small Brier score in the three-month periods}}
		\label{fig:Brierc}
	\end{subfigure}
	\quad
	\begin{subfigure}[b]{0.48\textwidth}
		\centering 
		\includegraphics[width=\textwidth]{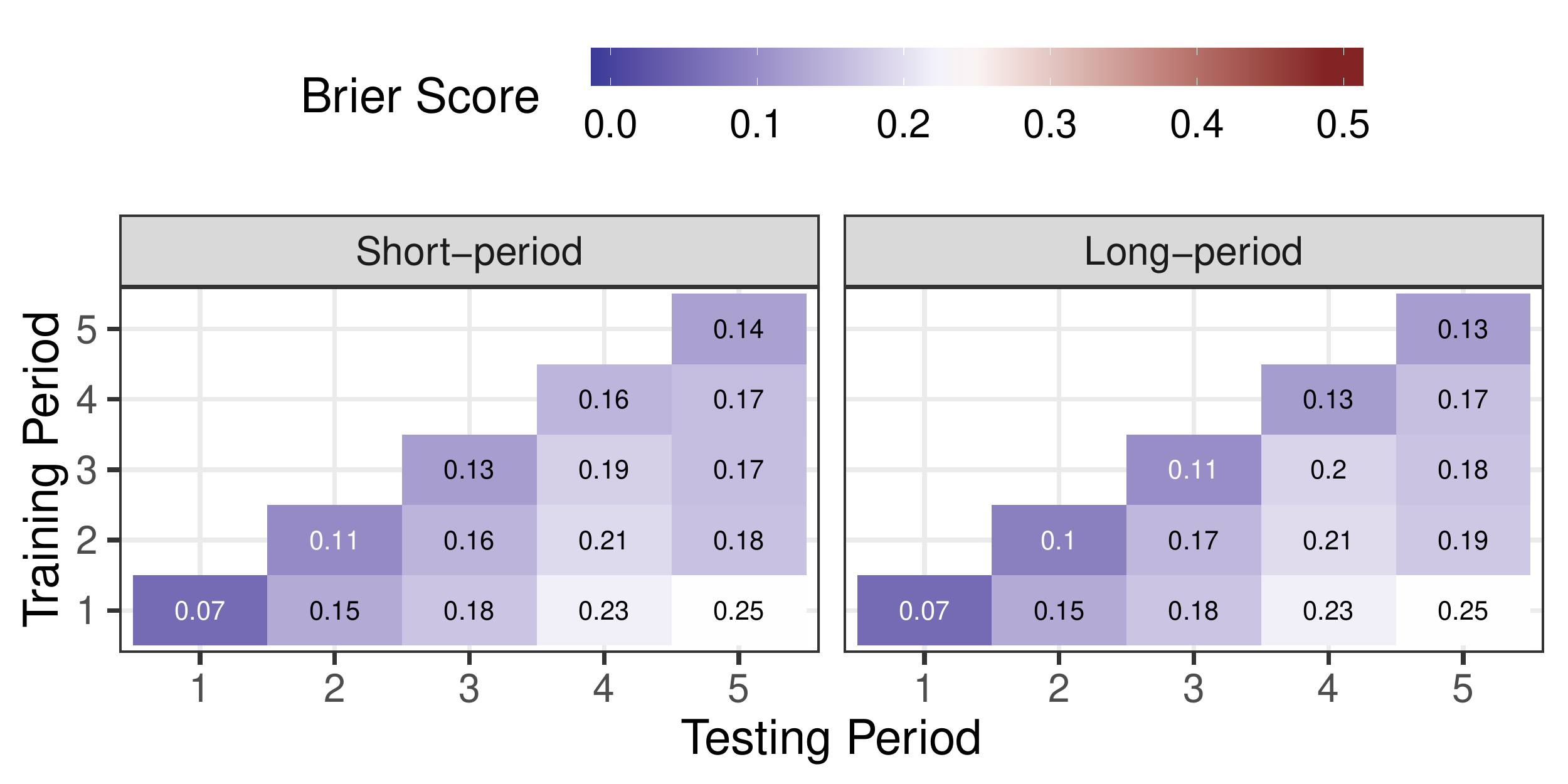}
		\caption[]%
		{{\small Brier score in the six-month periods}}
		\label{fig:Brierd}
	\end{subfigure}
	\caption[ The predictive performance of JIT models as the studied system age.
	]
	{\small The predictive performance of JIT models as the studied system age.} 
	\label{fig:predictiveperformance}
\end{figure}

\textbf{Results}
Figure~\ref{fig:predictiveperformance} shows heat-maps with the trend in AUC and Brier performance scores for each period tested for our short and long period JIT models.
The shade of the box indicates the performance value (from 0 to 1): Blue colors stand for strong performance, white colors for random guessing performance, and red colors for weak performance.

The columns of Figure~\ref{fig:AUCa} show that the values tend to improve as the training period increases.
For instance, column 4 of the long period model has 0.66 of AUC score when the model was trained using period 1.
However, the AUC score is 0.71, an improvement of 5\% points, when it was trained using period 3.
All in all, the long period model in Figure~\ref{fig:AUCa} presents a steady AUC score improvement of 5-9\% points when we trained the models using the most recent data instead of data from period 1. While the short period model presents a AUC score improvement of 6-10\% points.
The columns in Figure~\ref{fig:Brierc} also show a rise in Brier scores of 1-5\% points for the long period and 1-4\% points for the short period.
The six-month period models have almost the same performance, Figure~\ref{fig:AUCb} and Figure~\ref{fig:Brierd} show AUC and Brier improvements that reach 7-8\% and 3-8\% points for the long period, respectively.

The columns of Figure~\ref{fig:AUCa} show as well an improving trend in AUC scores that is more stable in long than in short period models.
For instance, columns 5, 6, and 7 show that the AUC improvement gained by adding the most recent period to the long period is 2 percentage point in column 5 (0.72 and 0.74 for training periods 3 and 4), 1 in column 6 (0.70 and 0.71 for training periods 4 and 5) and 0 in column 7 (0.71 and 0.71 for training periods 5 and 6).
While the improvement gained by adding the most recent period to periods 5, 6, and 7 in the short period models is 0, 2 and -1 respectively.
Figure~\ref{fig:AUCb} shows a similar tendency for six-month periods.
Figure~\ref{fig:Brierc} and Figure~\ref{fig:Brierd} indicate that the improving trend in the Brier score is stable in both, short and long period models.

When comparing these results with Mc\&K's paper, we noticed a considerable increase in the blue shades, which points out that our models perform stronger in terms of AUC scores. For three-month periods, JIT models without extrinsic bugs improved the AUC score from 1-16\% AUC points for testing periods 3-9 in the short and long periods. 
This improvement is for example noticeable in testing periods 3, 4, and 5 with training periods 1 and 2. While Mc\&K's models obtain almost the performance of a random guess, our models obtain an AUC score improvement of 6-16\% points for both short and long periods. 

Furthermore, after removing extrinsic bugs, our JIT models increase their stability by reducing two points in the short and long period (after period 5), while Mc\&K's models obtain a stability of -5.2\% and -1.2\% points for the short and long period, respectively; our models obtain -2.1\% and 0.2\% points for both periods respectively.

\begin{figure*}
	\centering
	\begin{subfigure}[b]{0.48\textwidth}
		\centering
		\includegraphics[width=\textwidth]{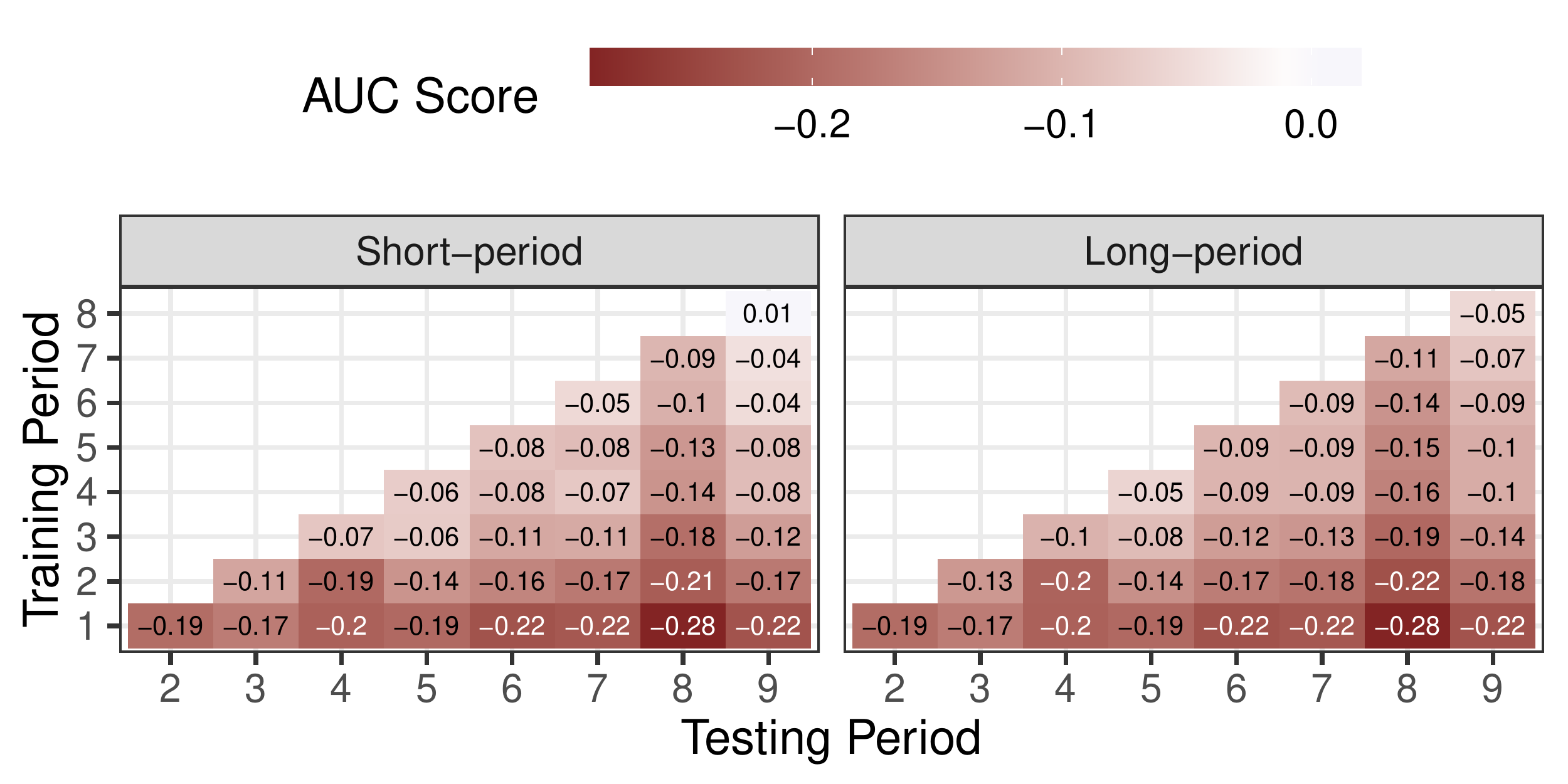}
		\caption[]%
		{{\small AUC in the three-month periods}}
		\label{fig:AUCDeltaa}
	\end{subfigure}
	\hfill
	\begin{subfigure}[b]{0.48\textwidth}
		\centering 
		\includegraphics[width=\textwidth]{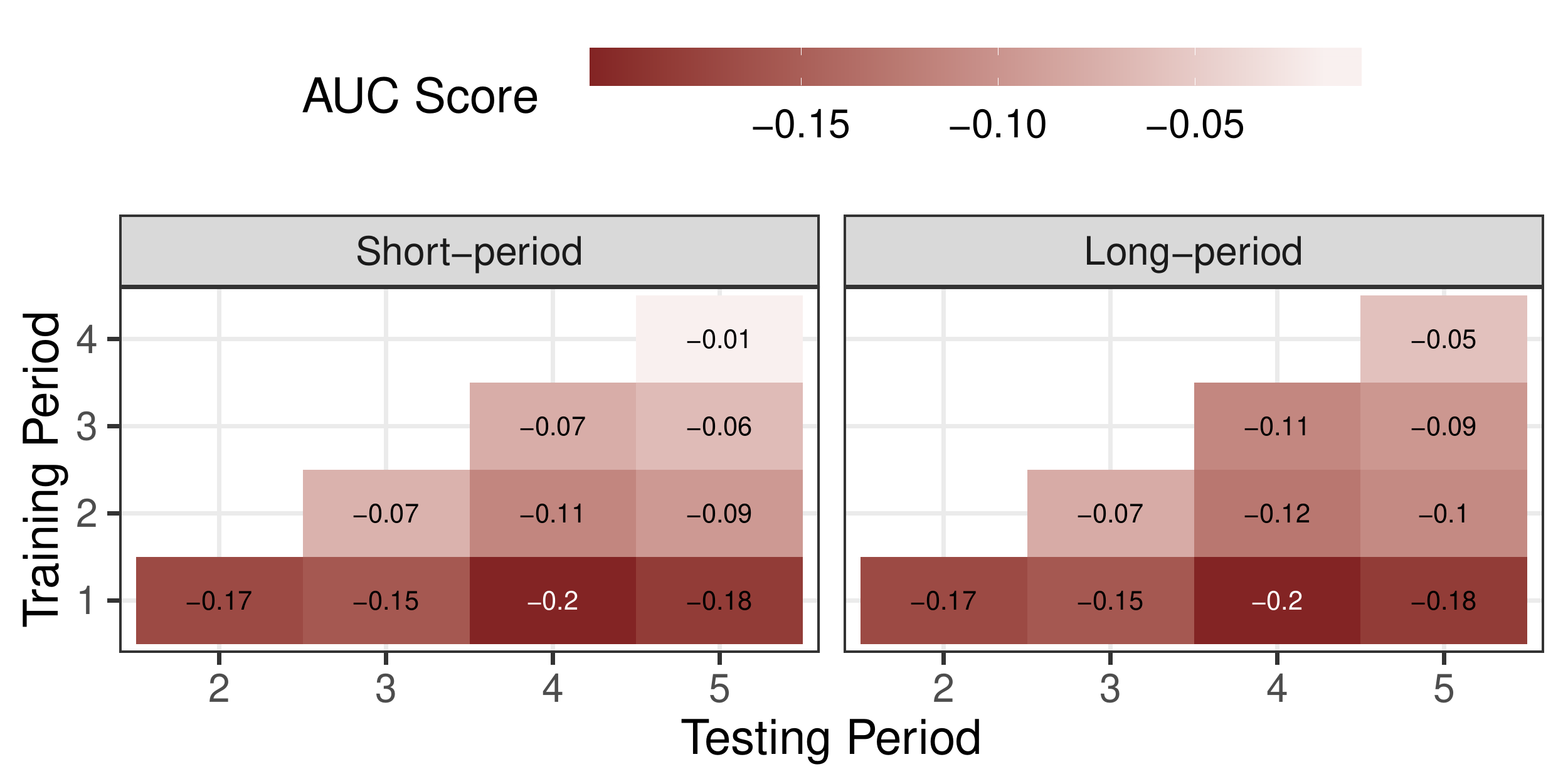}
		\caption[]%
		{{\small AUC in the six-month periods}}
		\label{fig:AUCDeltab}
	\end{subfigure}
	\vskip\baselineskip
	\begin{subfigure}[b]{0.48\textwidth}
		\centering 
		\includegraphics[width=\textwidth]{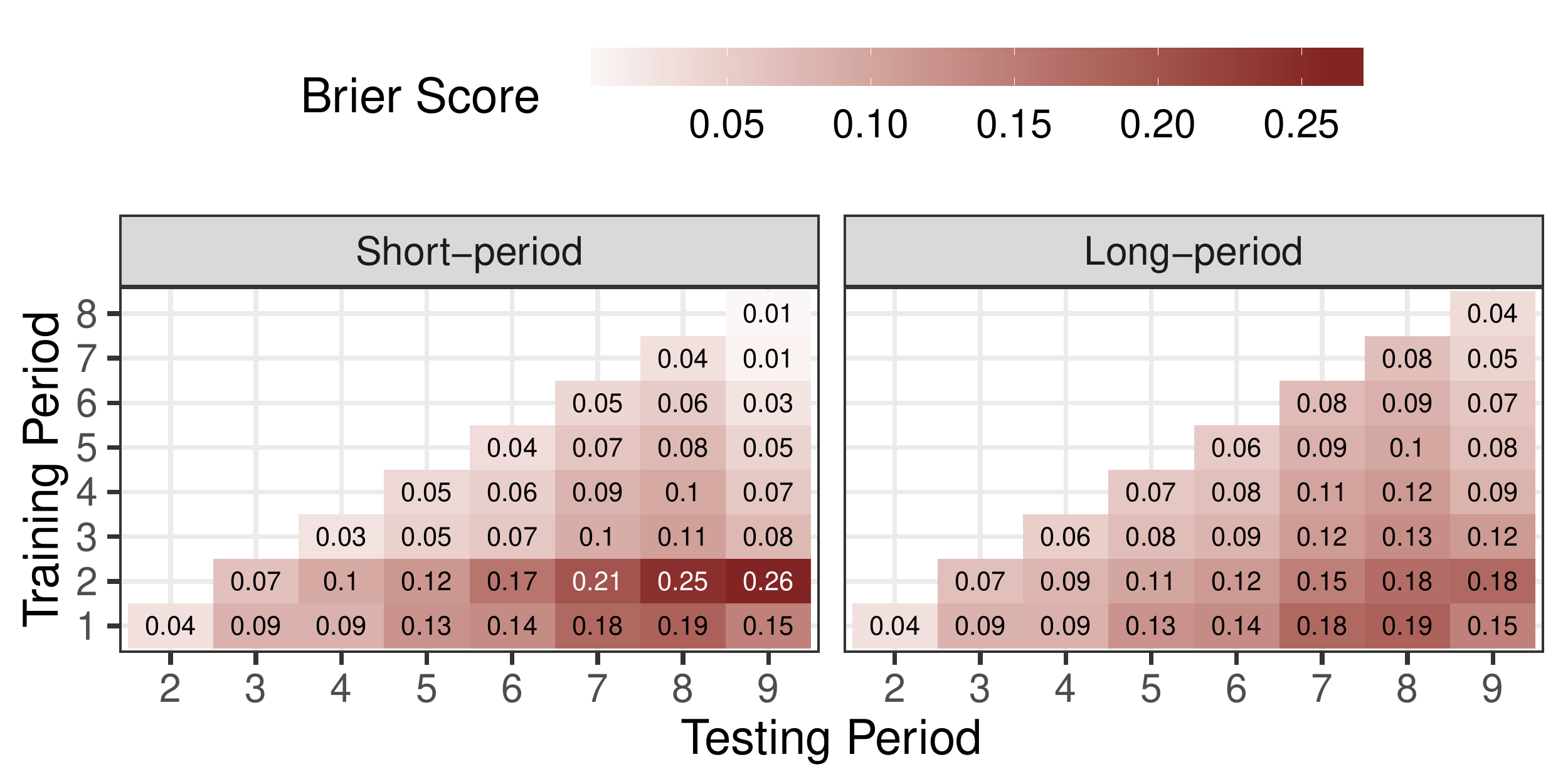}
		\caption[]%
		{{\small Brier score in the three-month periods}}
		\label{fig:BrierDeltac}
	\end{subfigure}
	\quad
	\begin{subfigure}[b]{0.48\textwidth}
		\centering 
		\includegraphics[width=\textwidth]{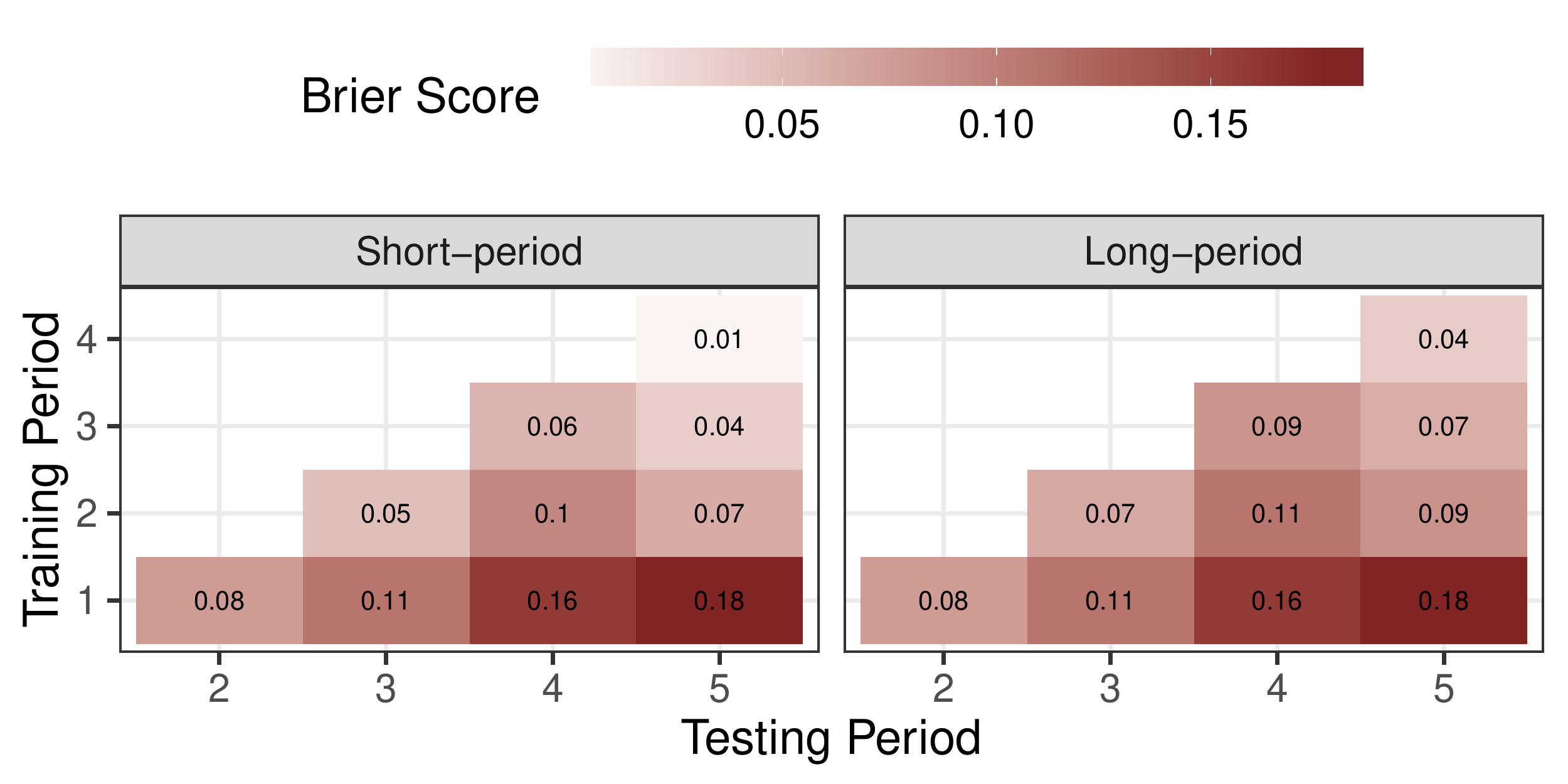}
		\caption[]%
		{{\small Brier score in the six-month periods}}
		\label{fig:Brierdeltad}
	\end{subfigure}
	\caption[ The delta in the estimate performance of JIT models as the studied system age.
	]
	{\small The delta in the estimate performance of JIT models as the studied system age.} 
	\label{fig:deltaperformance}
\end{figure*}

Figure~\ref{fig:deltaperformance} shows a heat-maps of the difference in AUC and Brier performance scores between training and testing periods over time of our short and long period JIT models.
The shade of the box offers information about the difference: blue colors stand for improvements performance, white colors for unchanged performance, and red colors for drops in performance in the testing period.\\

The analysis of the rows in Figure~\ref{fig:deltaperformance} offers evidence that our models lose predictive power after 12 months of being trained.
Figure~\ref{fig:AUCDeltaa} and Figure~\ref{fig:AUCDeltab} show that our short and long period models lose 8-19\% and 10-19\% AUC points 12 months after being trained (i.e., testing period = training period + 4) respectively.
Both figures show that after 12 months there is (often) a drop in the AUC. At the same time we can observe a boost in Brier scores (see Figures~\ref{fig:BrierDeltac} and~\ref{fig:Brierdeltad} respectively).

Thus, similar to Mc\&K's results, we lose predictive power in our JIT models after one year of being trained.
However, our models lose less amount of predictive power in each period when using testing periods 1, 2, 3, and 4. Also, AUC scores are more stable after removing extrinsic bugs. For example, Mc\&K's models lose 3-34\% AUC points in short period models after one year, while our models only lose 8-19\% AUC points. Thus, our models lose 15\% AUC points less and gained stability up to 20\% AUC points.

To observe the predictive power of long and short period JIT models, we focus on the data from period 2 and later since the AUC and Brier values of period 1 are identical in both periods. This is because there is no additional data added when training the long period model.
The rows of Figure~\ref{fig:AUCDeltaa} and Figure~\ref{fig:AUCDeltab} show that the short period models of periods 3 and later retain more predictive power than their long period counterparts in terms of AUC, i.e., the drop in the AUC values is smaller since these values are close to 0. Figure~\ref{fig:AUCDeltaa} shows that when the long period model is trained using period 3, it drops 10\% AUC points when it is tested in period 4, while it only drops 7\% AUC points in the short period model under the same circumstances.

Figure~\ref{fig:AUCDeltab} offers evidence that with six-month periods, both models retain similar predictive power; period 5 drops 8\% AUC points in both models. 
Figures~\ref{fig:BrierDeltac} and~\ref{fig:Brierdeltad} show that there is also an improvement in the retention of Brier score in short period models.
Furthermore, Figure~\ref{fig:deltaperformance} indicates that short period JIT models retain more predictive power than long periods.


\vspace{0.2cm}
\fbox{\begin{minipage}{23em}
		\textbf{Answer to RQ2:} 
		When removing extrinsic bugs, JIT models obtain better performance in terms of AUC (up to 16\% AUC points). 
		Models that only consider intrinsic bugs also lose predictive power 12 months after being trained, but they lose up to 15\% AUC points less and are up to 20\% AUC points more stable. 
\end{minipage}}

\subsection{\textbf{RQ3: How does the relationship between code change properties and the likelihood of BICs evolve in terms of time when extrinsic bugs are removed? }}

\textbf{Approach}
To answer this question, we followed Mc\&K's approach~\cite{mcintosh2018fix} and computed the normalized Wald $\chi\textsuperscript{2}$ importance score (see Section~\ref{modelanalysis}) for each family of code change properties, and for short and long period JIT models.
Furthermore, we computed the $\rho$-values associated with these scores.

\begin{figure}
	\centering
	\begin{subfigure}[b]{0.48\textwidth}
		\centering
		\includegraphics[width=\textwidth]{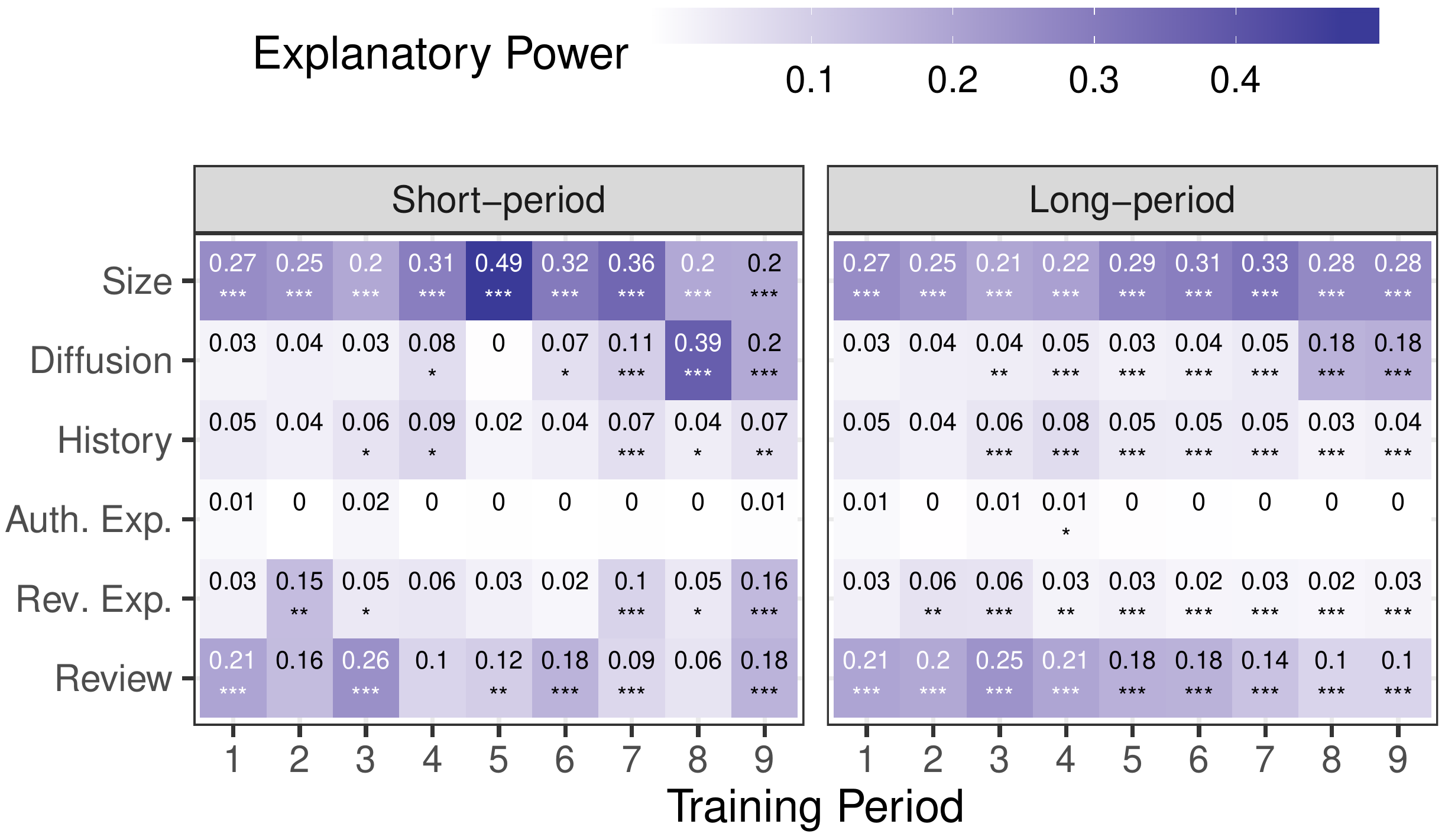}
		\caption[]%
		{{\small Three-month periods}}
		\label{fig:importance4}
	\end{subfigure}
	\begin{subfigure}[b]{0.48\textwidth}
		\centering 
		\includegraphics[width=\textwidth]{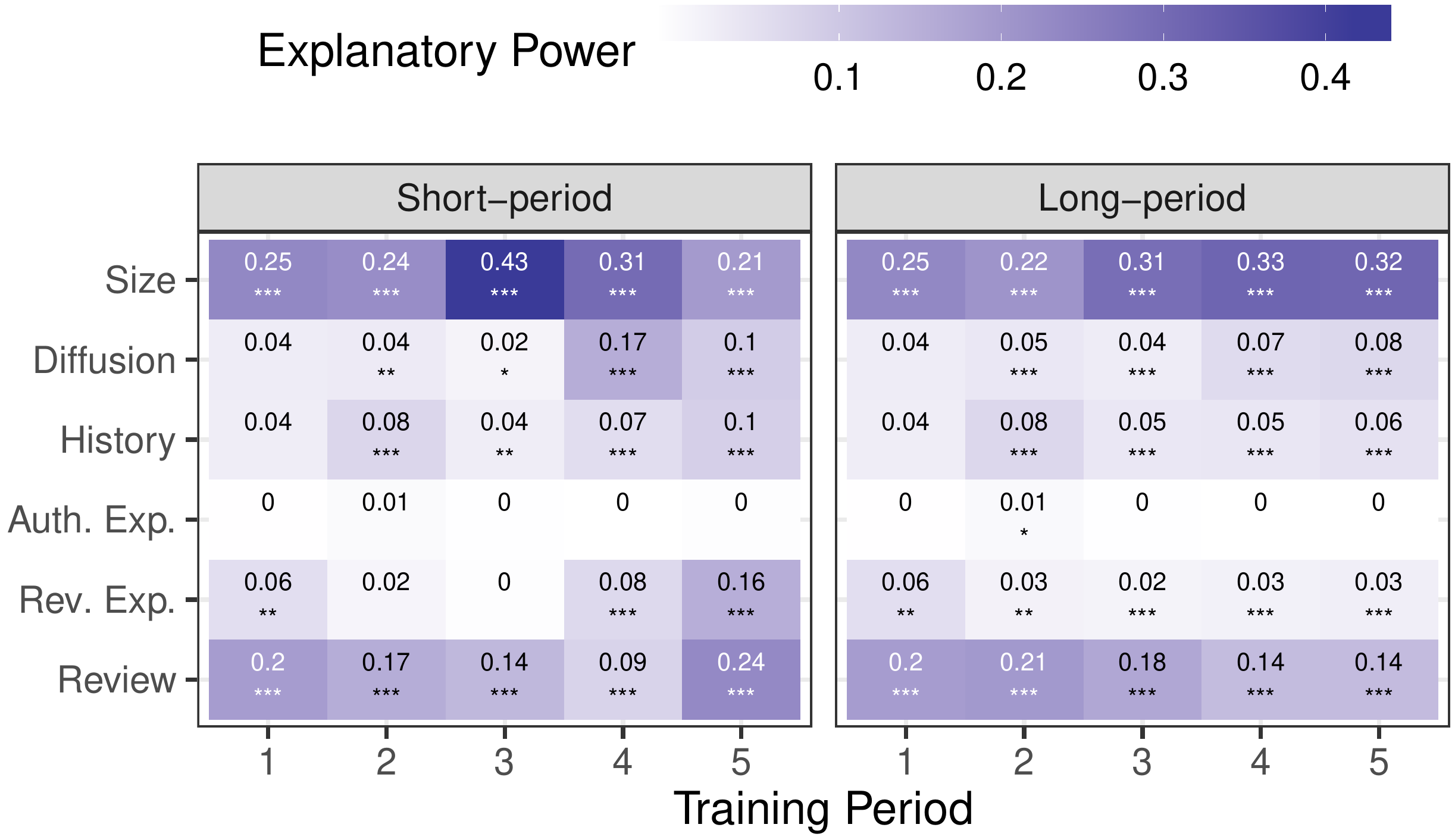}
		\caption[]%
		{{\small Six-month periods}}
		\label{fig:importance2}
	\end{subfigure}
	\caption[Evolution of the importance scores of the six studied families of code change properties over time.
	Shades indicates magnitude while asterisks indicate significance according to Wald $\chi\textsuperscript{2}$ test, where:\textsuperscript{*} $\rho< 0.05$; \textsuperscript{**}$\rho< 0.01$; \textsuperscript{***}$\rho< 0.001$.
	]
	{\small Evolution of the importance scores of the six studied families of code change properties over time.
		Shades indicates magnitude while asterisks indicate significance according to Wald $\chi\textsuperscript{2}$ test, where:\textsuperscript{*} $\rho< 0.05$; \textsuperscript{**}$\rho< 0.01$; \textsuperscript{***}$\rho< 0.001$.
	} 
	\label{fig:importance}
\end{figure}

\textbf{Results}
Figure~\ref{fig:importance} offers a series of heat-maps with the importance score of the six code change property families.
The darker the shade of the box, the more important the family is to our model.

Figure~\ref{fig:importance4} shows that in both short and long period models of three-month periods, the families of code changes that contribute the most are \emph{Size}, \emph{Diffusion}, and \emph{Review} for the last periods. 
\emph{Size} accounts for 20-49\%, \emph{Diffusion} for 0-39\%, and \emph{Review} for 6-26\% of the explanatory power in the short period models. In the long period models, \emph{Size} accounts for 21-33\%, \emph{Diffusion} for 3-18\%, and \emph{Review} for 10-25\% of the explanatory power.

The six-month period models present similar results. Figure~\ref{fig:importance2} shows that the \emph{Size} and \emph{Review} families account for more of the explanatory power in both short and long period models. \emph{Size} accounts for 21-43\% and 22-33\%, and \emph{Review} for 9-24\% and 14-21\%. 

For the six-month periods models, Figure~\ref{fig:importance} shows that the \emph{Size} family is the top contributor in all periods of both short and long period models. 
For the three-month periods models, Figure~\ref{fig:importance} shows that, in both short and long period models, the \emph{Size} family is the top contributor in 8 out of 9 periods. The \emph{Review} family is the top contributor in the remaining periods.

The contributed explanatory power of the \emph{Size} family is statistically significant ($\rho<0.01$, $\rho<0.001$) in all of the periods for our long and short period models in the three-month and six-month periods.
The \emph{Review} family's explanatory power is also statistically significant ($\rho<0.01$, $\rho<0.001$) in all of the periods in the long period model of the three-month periods and for both models of the six-month periods.
However, in the short period models of the three-month periods, the \emph{Review} family's explanatory power is statistically significant 
only in 6 out of 9 periods.

Compared to Mc\&K's paper, when removing extrinsic bugs in both short period models of three- and six-month periods, the explanatory power of the \emph{Size} family increases 
from 3-37\% to 20-49\%, and from 16-25\% to 24-43\%. However, the explanatory power of the \emph{Review} family decreases considerably from 2-59\% to 6-26\%, and from 8-38\% to 9-24\%. In both long period models of three- and six-month periods, when removing extrinsic bugs, the explanatory power of the \emph{Size} family also increases from 11-37\% to 21-33\%, and from 15-19\% to 22-33\%, but the explanatory power of the \emph{Review} family decreases considerably from 15-43\% to 10-25\%, and from 24-37\% to 14-21\%.
This may indicate that extrinsic bugs have different characteristics affecting the \emph{Review} family.
Moreover, the statistical significance power of the \emph{Review} and \emph{Size} families also increases when removing extrinsic bugs.
Our models increase the number of periods with statistical significance of the \emph{Diffusion} family in both long and short periods of the three-months periods and six-months periods.

Furthermore, our models increase the number of the significant periods in \emph{Diffusion} for both long and short six-month and three-month period models. However, the number of significant periods of the \emph{History} family decreases for both long and short three- and six-month period models.

Finally, fluctuations of the properties of BICs are more stable in our JIT models. This suggests that although properties of intrinsic bugs tend to evolve as projects age, the properties of extrinsic bugs fluctuate more drastically from period to period.

\vspace{0.1cm}
\fbox{\begin{minipage}{23em}
		\textbf{Answer to RQ3:} 
		When removing extrinsic bugs, the importance of the \emph{Size} family increases (up to 18\% AUC points), but the importance of the \emph{Review} family decreases (up to 36\% AUC points). 
		Furthermore, the importance of most families of code changes are more stable through periods, suggesting that the properties of BICs tend to evolve less drastically with the project over time.
\end{minipage}}

\subsection{\textbf{RQ4: How accurately do current importance scores of code change properties represent future ones when extrinsic bugs are removed?}}
\textbf{Approach}
As Mc\&K's paper, we used the Family Importance Score~\emph{(FIS}) metric to study the stability of the importance scores of each family of code change properties. FIS(f,n) is the jointly tested model terms for all metrics belonging to a family $f$ in the model of period $n$. 

These periods can be the training periods which are represented by $i$, or the testing periods --or future periods-- which are represented by $j$. Thus, for each one of the JIT models (short and long period) and for each family $f$, we computed the differences between the importance scores of each family in the training periods $i$ and future periods $j$ using \emph{FISDiff($f,i,j$)} = \emph{FIS($f,i$)} - \emph{FIS($f,j$)}.



When the difference between the importance scores of a family $f$ in periods $i$ and $j$ is higher than 0, this family has a larger importance in period $i$ (training) than in period $j$ (future).
In such cases, the JIT model (trained using period $i$) \emph{overestimates} the future importance of family $f$.
On the contrary, when that difference is lower than 0, it indicates that family $f$ has smaller importance in period $i$ (training) than in period $j$ (future).
If this occurs, the JIT model (trained using period $i$) \emph{underestimates} the future importance of family $f$.

When the model overestimates the future importance of a family $f$, the impact of that family at the end of the period might be smaller than anticipated. On the other hand, when the model underestimates the future importance of a family $f$, the impact of that family at the end of the period might be bigger than anticipated. Software Quality Assurance (SQA) teams can use these importance scores to estimate quality improvements for future periods.

\begin{figure*}
	\centering
	\begin{subfigure}[b]{0.65\textwidth}
		\centering
		\includegraphics[width=\textwidth]{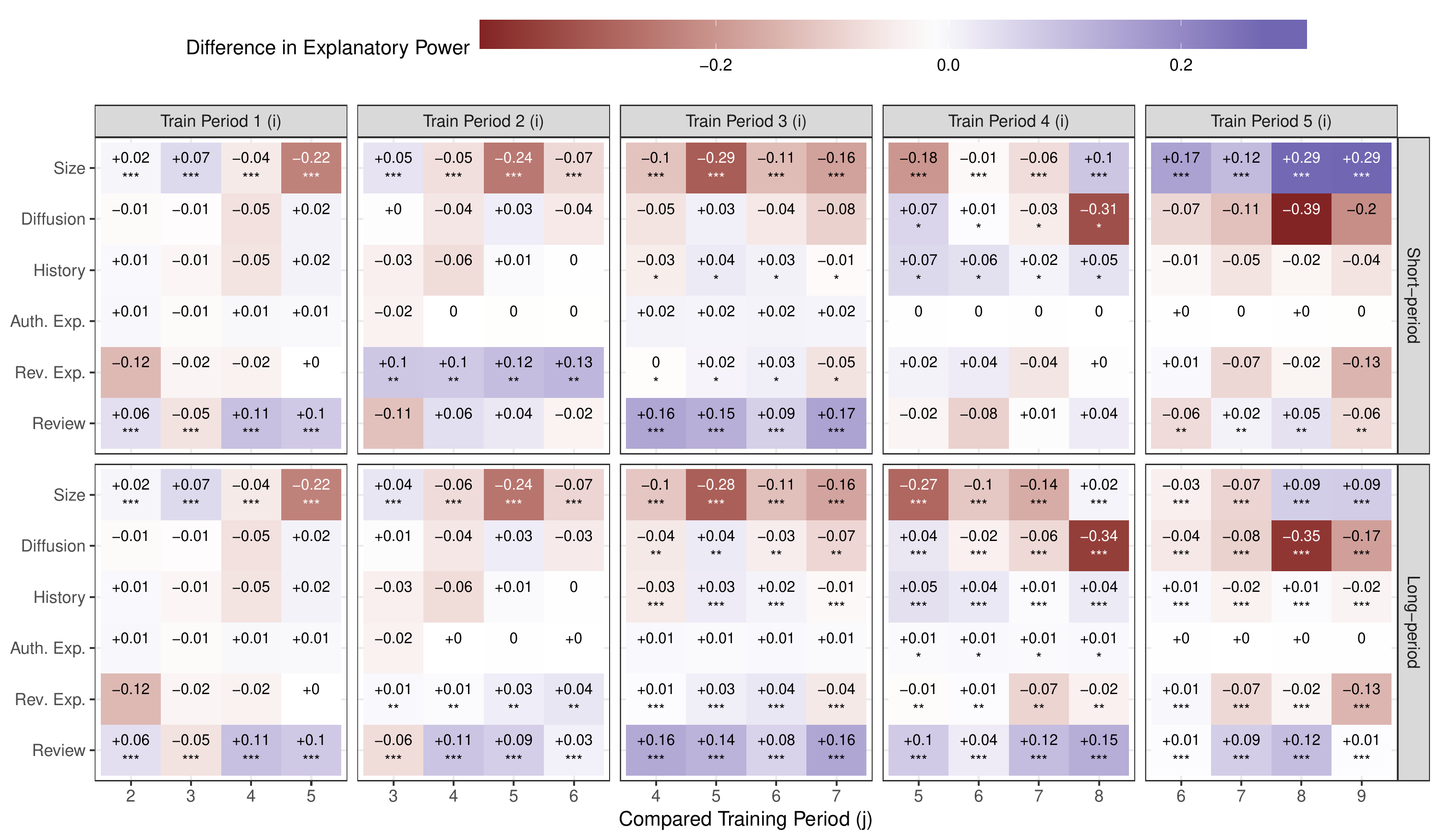}
		\caption[]%
		{{\small Three-month periods}}
		\label{fig:stability4}
	\end{subfigure}
	\hfill
	\begin{subfigure}[b]{0.30\textwidth}
		\centering 
		\includegraphics[width=\textwidth]{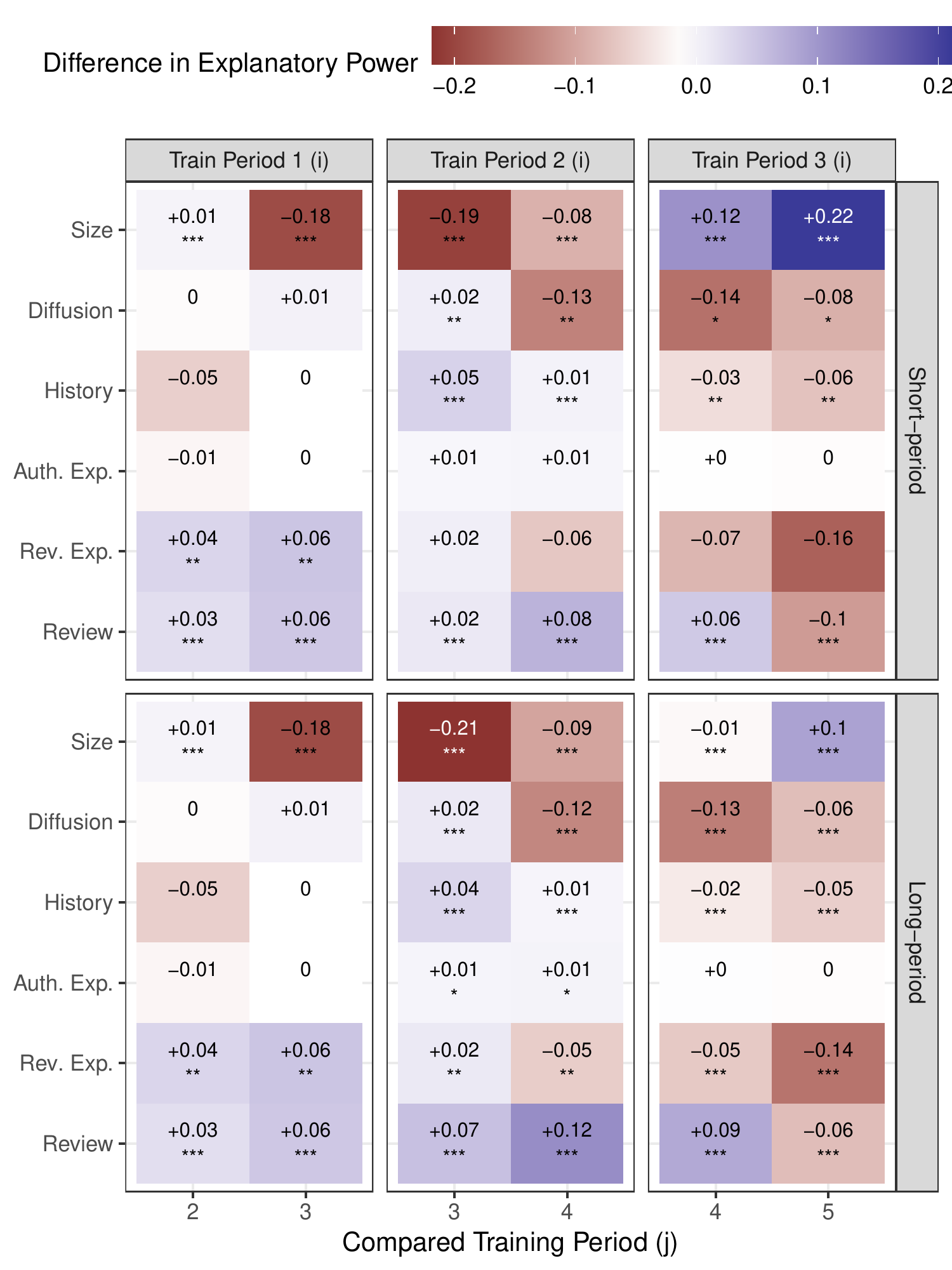}
		\caption[]%
		{{\small Six-month periods}}
		\label{fig:stability2}
	\end{subfigure}
	\caption[ The stability of the importance scores of the studied families of code change properties (FISDiff(\emph{f,i,j})).
	]
	{\small The stability of the importance scores of the studied families of code change properties (FISDiff(\emph{f,i,j})).
	} 
	\label{fig:stability}
\end{figure*}

\textbf{Results}
Figure~\ref{fig:stability} presents a series of heat-maps with the differences between the importance score in period $i$ and $j$ for each of the six code change property families.
Furthermore, each cell reports the statistical significance of the importance score.


In the three-month period models, Figure~\ref{fig:importance4} shows that the \emph{Size} family spikes in period 5 with a score of 0.49.
Training periods 1, 2, 3, and 4 in Figure~\ref{fig:stability4} show that the importance of \emph{Size} is underestimated by 22\%, 24\%, 29\%, and 18\% AUC points respectively for testing period 5 in short periods.
In the long period models, the underestimation of the importance of \emph{Size} for testing period 5 has similar values.
When period 5 becomes the training period in Figure~\ref{fig:stability4}, the importance of the \emph{Size} family is overestimated in the short period model by up to 29\% AUC points. However, in the long period model, the maximum overestimation is significantly smaller: 9\% AUC points.

The short period models of Figure~\ref{fig:stability4} shows several fluctuations in the importance score of each family over the periods.
In the six-month period models, Figure~\ref{fig:stability2} shows the same trend for the \emph{Size} family but with less severe overestimation or underestimation.

Thus, similar to Mc\&K's paper, the importance of the \emph{Size} family is underestimated while the \emph{Review} family is overestimated when removing extrinsic bugs, especially in training periods 1, 2, and 3. 
However, we found that either long or short period models perform similar. The fluctuations in importance in long period models are not smoother than the fluctuations in short periods. 

\vspace{0.1cm}
\fbox{\begin{minipage}{23em}
		\textbf{Answer to RQ4:} 
		When removing extrinsic bugs, long-period models do not outperform short periods when analyzing the stability of the importance scores. Larger amounts of training data will not smooth the impact or fluctuations between periods.
\end{minipage}}

\subsection{ \textbf{RQ5: How do mislabeled bugs affect JIT models?}}
\label{RQ5}
\textbf{Approach}
Although we manually identified mislabeled bugs, we decided to include them into the dataset that fed JIT bug prediction models, i.e., we just removed extrinsic bugs in answers RQ2-RQ4. The reason for doing this is because Tantithamthavorn~\emph{et al.} recently found that mislabeled bugs do not have much impact in defect prediction when analyzing whether a file will be buggy or not~\cite{tantithamthavorn2015impact}, so we expected it to be the same for JIT models.
With RQ5, we want to evaluate if this is true.

We created a ground truth dataset by removing extrinsic and mislabeled bugs from Mc\&K's dataset. Since this dataset only contains intrinsic bugs, the most accurate JIT bug prediction models are to be obtained when using this dataset for training the models. Therefore, to study the impact of mislabeled bugs on JIT bug prediction models, we compared the results obtained after training JIT models using the ground truth dataset with the results obtained after training JIT models using intrinsic and mislabeled bugs.

Finally, to obtain a complete picture, we compared the results using the ground truth dataset (i.e., only intrinsic bugs) with the results of (1) intrinsic and mislabeled bugs, (2) intrinsic and extrinsic bugs, and (3) intrinsic, extrinsic, and mislabeled bugs (i.e., Mc\&K's results).

\textbf{Results}. While manually curating the dataset, we have identified 548 mislabeled bugs (29.1\%) and 212 extrinsic bugs (11\%) in Mc\&K's dataset. 
The percentage of mislabeled bugs is similar to the percentage reported in previous studies, ranging from 33\% to 40\%~\cite{herzig2013s,rodriguez2016bugtracking}. Furthermore, the percentage of extrinsic bugs is also similar to the one reported in previous studies (9-21\%)~\cite{rodriguez2019how}. 

Extrinsic bugs were removed in RQ2 and RQ3. To obtain the ground truth dataset, we removed the 548 mislabeled bugs from our dataset. Since mislabeled bugs are not bugs, they do not have a BIC. But, they have a BFC. So, we removed the link between issueIDs classified as mislabeled bugs and their BICs, and we trained again the JIT models, this time using this new dataset (i.e., using the ground truth dataset, composed of 1,120 issues and 1,571 BICs). 

We followed the procedures described in RQ2 and RQ3 to analyze the most accurate performance that JIT models can have using the ground truth dataset. We compared these results with (1) the performance of JIT models when mislabeled bugs are included in the dataset; (2) the performance of JIT models when extrinsic bugs are included in the dataset; and (3) the performance of JIT models when mislabeled and extrinsic bugs are included in the dataset. We will report the results of this RQ in textual form, due to space constraints. All figures corresponding to the ones in RQ2-RQ4 for the scenarios under study in RQ5 can be found in the online appendix\footnote{\url{http://gemarodri.github.io/2019-Study-of-Extrinsic-Bugs/}}. 

Table~\ref{comparison} shows the delta comparison between the ideal results (only intrinsic bugs) and the results of the different JIT models implemented for this RQ. We obtain a complete picture of how extrinsic bugs and mislabeled bugs affect the performance of JIT bug prediction models. A score of 0 in the table means that for that particular case, the JIT model performs as good as the ideal JIT model.

\textbf{Training JIT models with intrinsic and mislabeled bugs}: when the datasets contain intrinsic bugs and mislabeled bugs, the performance of the models decrease up to 4\% AUC points for both short and long periods of the three month period models. Furthermore, the performance also decreases 2\% AUC points for both short and long periods of the six month period models. These models are almost as stable as the models trained with only intrinsic bugs. 

The importance of the studied families differ from the ideal scenario. Although the importance of the \emph{Size} family is sightly overestimated, the importance of the \emph{Diffusion} and the \emph{History} families are overestimated up to 14\% AUC and 12\% AUC points for the three month short periods. Moreover, the \emph{History} family is underestimated up to 13\% AUC points for the three month long periods.

\textbf{Training JIT models with intrinsic and extrinsic bugs}: the performance of these models decreases up to 3\% AUC points for both short and long periods of the three month period models. However, the performance of both long periods of the three and six month period models increases up to 3\% AUC points. This means that these models are over-fitted. These models are as stable as the models trained with only intrinsic bugs for the three month long periods and the six month short periods. 

The importance of the \emph{Rev.Exp.} family is overestimated up to 12\% AUC points, but underestimated up to 10\% AUC points for the three month long periods and six month short periods, respectively. There are sightly no differences in the importance of the remaining families.

\textbf{Training JIT models with intrinsic, mislabeled, and extrinsic bugs}: the performance of these models increases up to 15\% AUC points and 9\% AUC points for both short and long periods of the three- and six-month period models respectively. Therefore, these models are over-fitted, which may cause a poor predictive performance.

The importance of the \emph{Size}, \emph{Diffusion}, and \emph{History} families is either overestimated or underestimated for the three and six month periods. The importance of the \emph{Review} family is overestimated up to 20\% AUC points for the three short period models.

\begin{table*}
	\renewcommand{\arraystretch}{1.2}
	\caption{Comparison of the results of the different JIT prediction models implemented in this study with respect to the ideal JIT bug prediction model (only intrinsic bugs). ``[Intrinsic+Mislabeled] Bugs" stands for the models after removing extrinsic bugs. ``[Intrinsic+Extrinsic] Bugs" stands for the models after removing mislabeled bugs. ``[Intrinsic+Mislabeled+ Extrinsic] Bugs" stands for McIntosh and Kamei's models~\cite{mcintosh2018fix}.
	}
	\label{comparison}
	\centering
	\begin{smaller}
		\begin{tabular}{l|p{0.88cm} p{0.88cm}|p{0.88cm} p{0.88cm}|p{0.88cm} p{0.88cm}|p{0.88cm} p{0.88cm}|p{0.88cm} p{0.88cm}|p{0.88cm} p{0.88cm}}
			\toprule
			&\multicolumn{4}{c|}{ \textbf{[Intrinsic + Mislabeled] Bugs}} &\multicolumn{4}{c|}{ \textbf{[Intrinsic + Extrinsic] Bugs}} &\multicolumn{4}{c}{\textbf{[Intrinsic + Mislabeled + Extrinsic] Bugs}}\\
			\hline
			\hline
			Issues & \multicolumn{4}{c|}{1,668} & \multicolumn{4}{c|}{1,332} & \multicolumn{4}{c}{1,880} \\
			Total BICs &   \multicolumn{4}{c|}{2,506} &  \multicolumn{4}{c|}{2,132}  &  \multicolumn{4}{c}{3,067} \\ 
			\hline \hline
			&\multicolumn{2}{c|}{ \textbf{3 months}} &\multicolumn{2}{c|}{ \textbf{6 months}}&\multicolumn{2}{c|}{ \textbf{3 months}} &\multicolumn{2}{c|}{ \textbf{6 months}}&\multicolumn{2}{c|}{ \textbf{3 months}} &\multicolumn{2}{c}{ \textbf{6 months}}\\
			\hline
			& \textbf{Short} & \textbf{Long} & \textbf{Short} & \textbf{Long}& \textbf{Short} & \textbf{Long}& \textbf{Short} & \textbf{Long}& \textbf{Short} & \textbf{Long}& \textbf{Short} & \textbf{Long} \\
			\hline
			$\Delta$ \% AUC &  [-4,3] &  [-4,1] & -2  & -2 & [-3,8]&[-3,3]&[-1,3]&[-1,3]&[-2,15]&[2,15]&[2,9]& [3,9]\\
			$\Delta$ Stability & [-1,1] & 1  & 1 &1 & [-3,-1]& \textbf{0}& \textbf{0}&2&[1,3]&-1&-1& -1\\
			$\Delta$ \% Size Fam. &[2,3]  & [1,-2]  &[4,5]  & 2 &  [-2,-1] &[-2,2]&[-2,5]&[-4,2]&[-10]&[-9,6]&[-13,-6]&[-5,12]\\
			$\Delta$ \% Diffusion Fam. & [14] & 4 & [1,10] &[1,2] & [1,2] &[2,3] &[2,4]&[-3,-2]&16&[-1,14]&10&2\\
			$\Delta$ \% History Fam. & [1,12]  & [-13,-3]  & [-3,-2]  & [-1,2]  & [2,3] &[-2,3] &[-4,2]&[1]&[2]&[-4,4]&[-4,6]&[-1,10]\\
			$\Delta$ \% Auth.Exp. Fam. & \textbf{0} &  \textbf{0} & \textbf{0} &1 &3 &1 &1&-2&3&3&1&2\\
			$\Delta$ \% Rev.Exp. Fam. & [-7,1] & -2  &-7   & -2  & -2 &[1,12] &[-5,1]&[-10,-2]&6&[-4,-3]&6&[3,4]\\
			$\Delta$ \% Review Fam. & [-3,-2]  & -10  & -9   & [1,-8]  & [3,7] &[-3,-1] &[-3]&[1,3]&[-2,20]&[5,8]&[1,3]&[2,11]\\
			\hline
			\bottomrule 
		\end{tabular}
	\end{smaller}
\end{table*}

\vspace{0.2cm}
\fbox{\begin{minipage}{24em}
		\textbf{Answer to RQ5:} 
Mislabeled bugs affect JIT models reducing their performance up to 4\% AUC points and underestimating the importance of the \emph{History} family. Extrinsic bugs affect JIT models reducing their performance up to 3\% AUC points and overestimating the importance of the \emph{Rev.Exp.} family. 
\end{minipage}}

\subsection{ \textbf{RQ6: Are the properties BFCs and BICs linked to extrinsic, intrinsic, and mislabeled bugs different?}}

\textbf{Approach}
During the manual classification, we found that Mc\&K's dataset not only contained extrinsic bugs, but also mislabeled bugs. Thus, to further understand whether code change properties are different among these categories, we analyzed the distributions and probability density of the six families of code change properties (see Table~\ref{taxonomy}) of (1) the commit identified as BFCs and (2) the commits identified as BICs. Notice that, although extrinsic bugs and mislabeled bugs cannot be linked to a BIC, in this RQ we analyze whether there are differences between the BICs linked to intrinsic bugs and those BICs (incorrectly) linked to extrinsic bugs and mislabeled bugs. 

We then compute whether the differences between these three groups were statistically significant across the six families of code change properties for BFCs and BICs. For that, we used the the Kruskal‐Wallis test~\cite{mckight2010kruskal}. This test is a non-parametric statistical test that assesses the differences among three or more independently sampled groups on a single, non‐normally distributed continuous variable\footnote{We found that the final dataset contained skewed data using the function \emph{skewness} with the \emph{e1071} package in R.}. 

Finally, we analyze how different these three groups are when they are paired in two groups i.e., Extrinsic-Intrinsic, Extrinsic-Mislabeled, and Intrinsic-Mislabeled. For that, we used the Wilcoxon Signed Rank test~\cite{whitley2002statistics} which is a non-parametric test that statistically compares the average of two dependent samples and assesses for significant differences.

\textbf{Results}
\textbf{a) Code Change Properties of BICs for intrinsic, extrinsic, and mislabeled bugs}

\begin{figure*}
	\centering
	\begin{subfigure}[b]{0.323\textwidth}
		\centering
		\includegraphics[width=\textwidth]{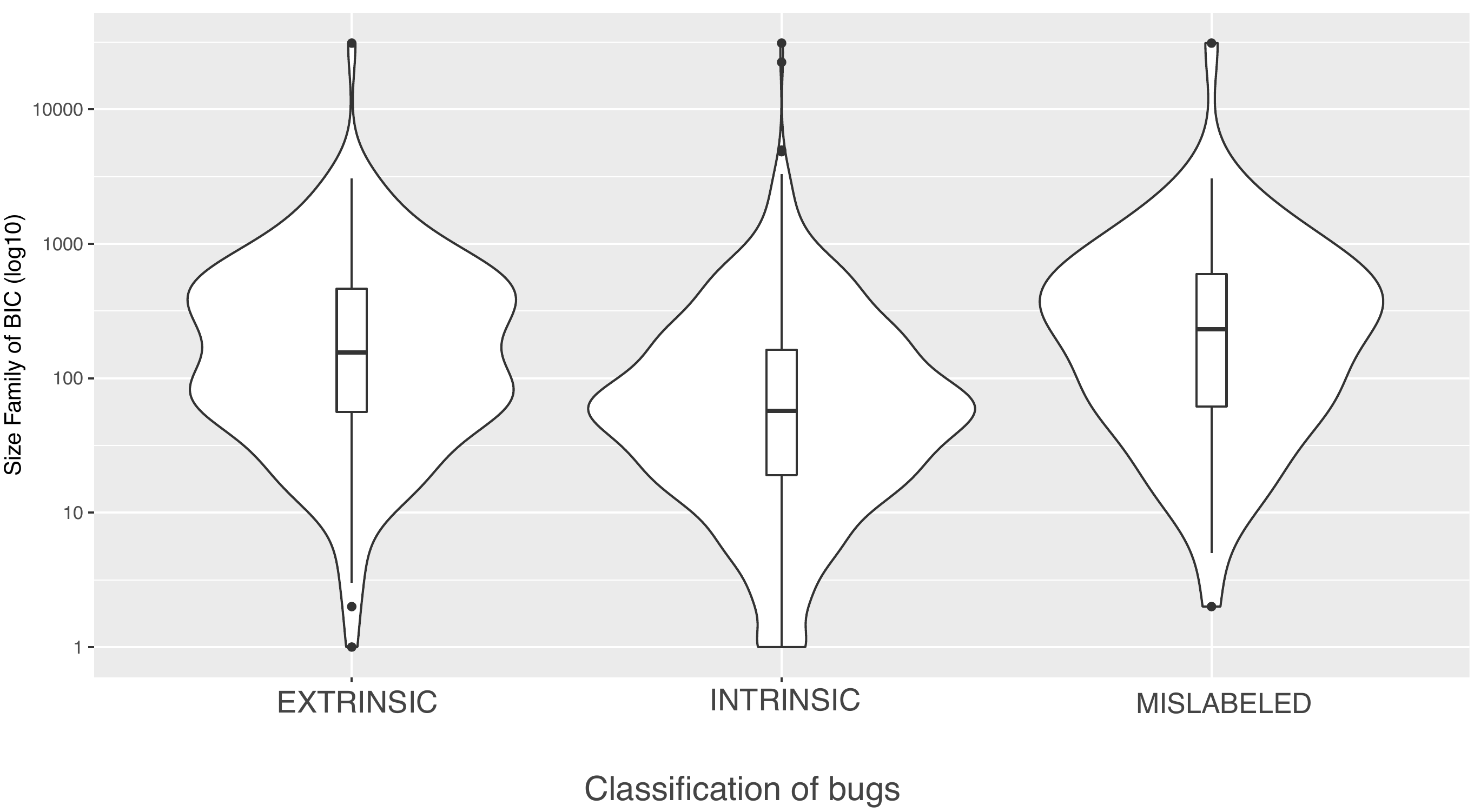}
		\caption[]%
		{{\small Size}}
		\label{fig:size}
	\end{subfigure}
	\hfill
	\begin{subfigure}[b]{0.323\textwidth}
		\centering
		\includegraphics[width=\textwidth]{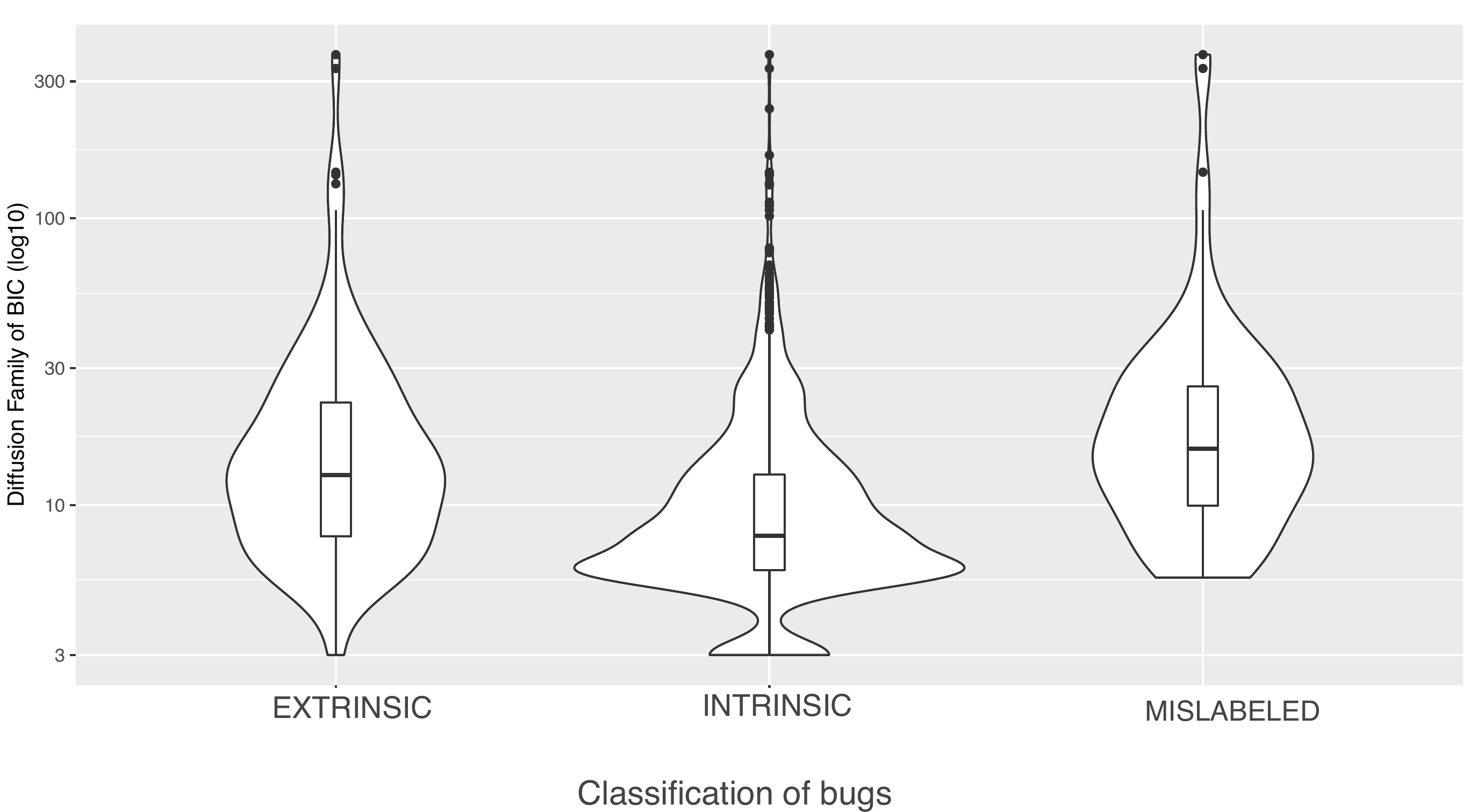}
		\caption[]%
		{{\small Diffusion}}
		\label{fig:diffusion}
	\end{subfigure}
	\hfill
	\begin{subfigure}[b]{0.323\textwidth}
		\centering 
		\includegraphics[width=\textwidth]{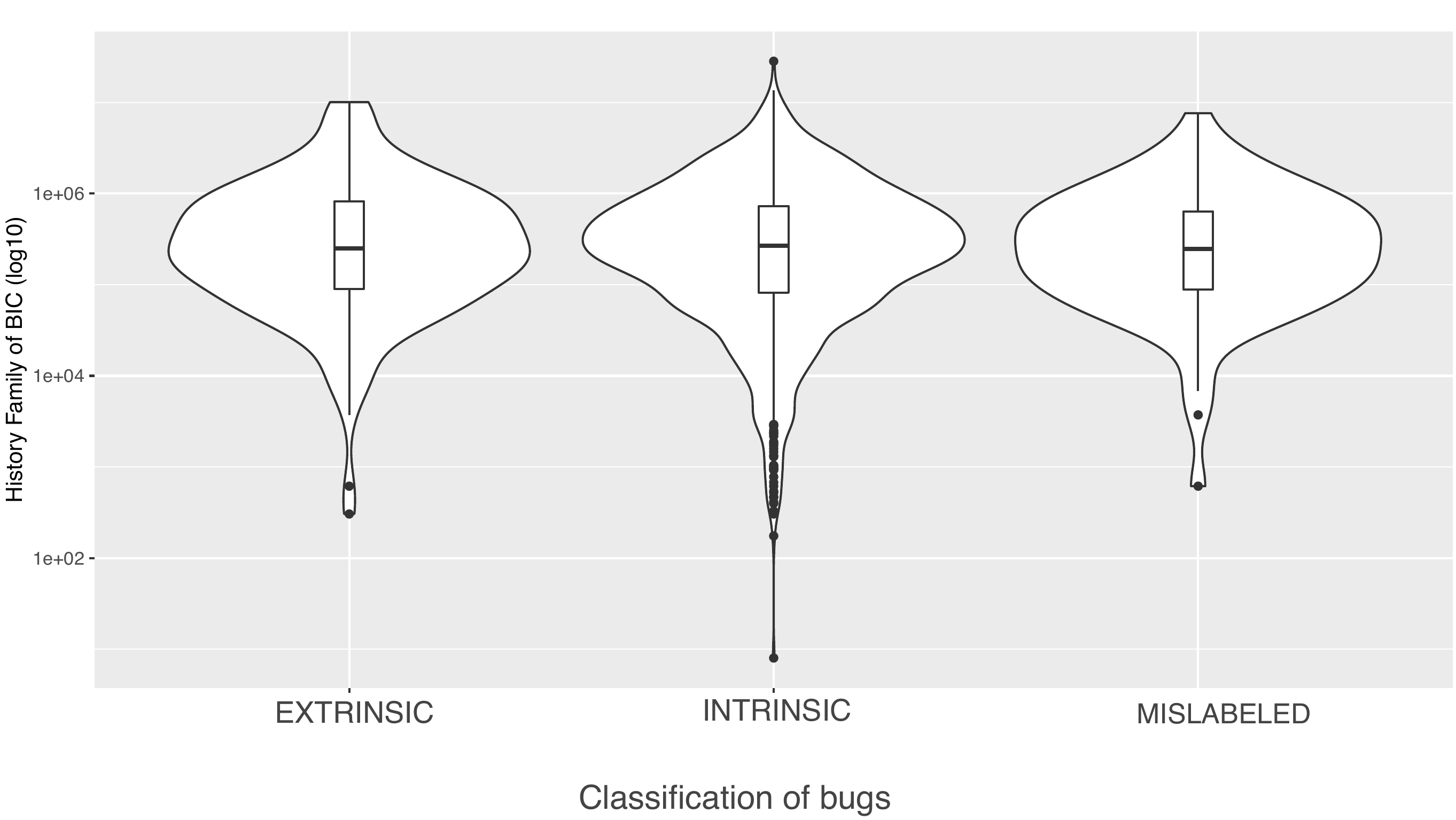}
		\caption[]%
		{{\small History}}
		\label{fig:history}
	\end{subfigure}
	\vskip\baselineskip
	\begin{subfigure}[b]{0.33\textwidth}
		\centering 
		\includegraphics[width=\textwidth]{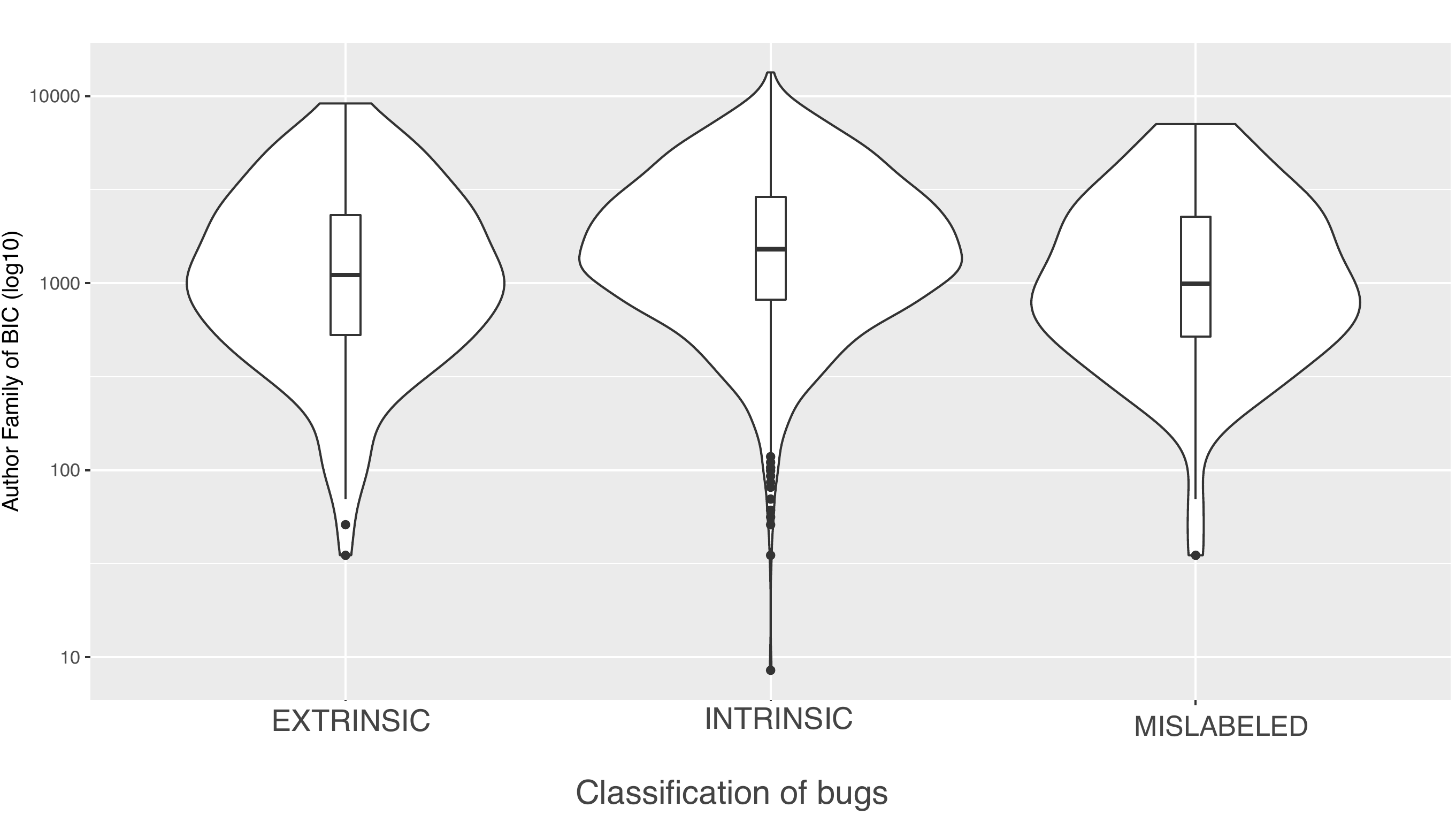}
		\caption[]%
		{{\small Author}}
		\label{fig:author}
	\end{subfigure}
	\hfill
	\begin{subfigure}[b]{0.33\textwidth} 
		\centering 
		\includegraphics[width=\textwidth]{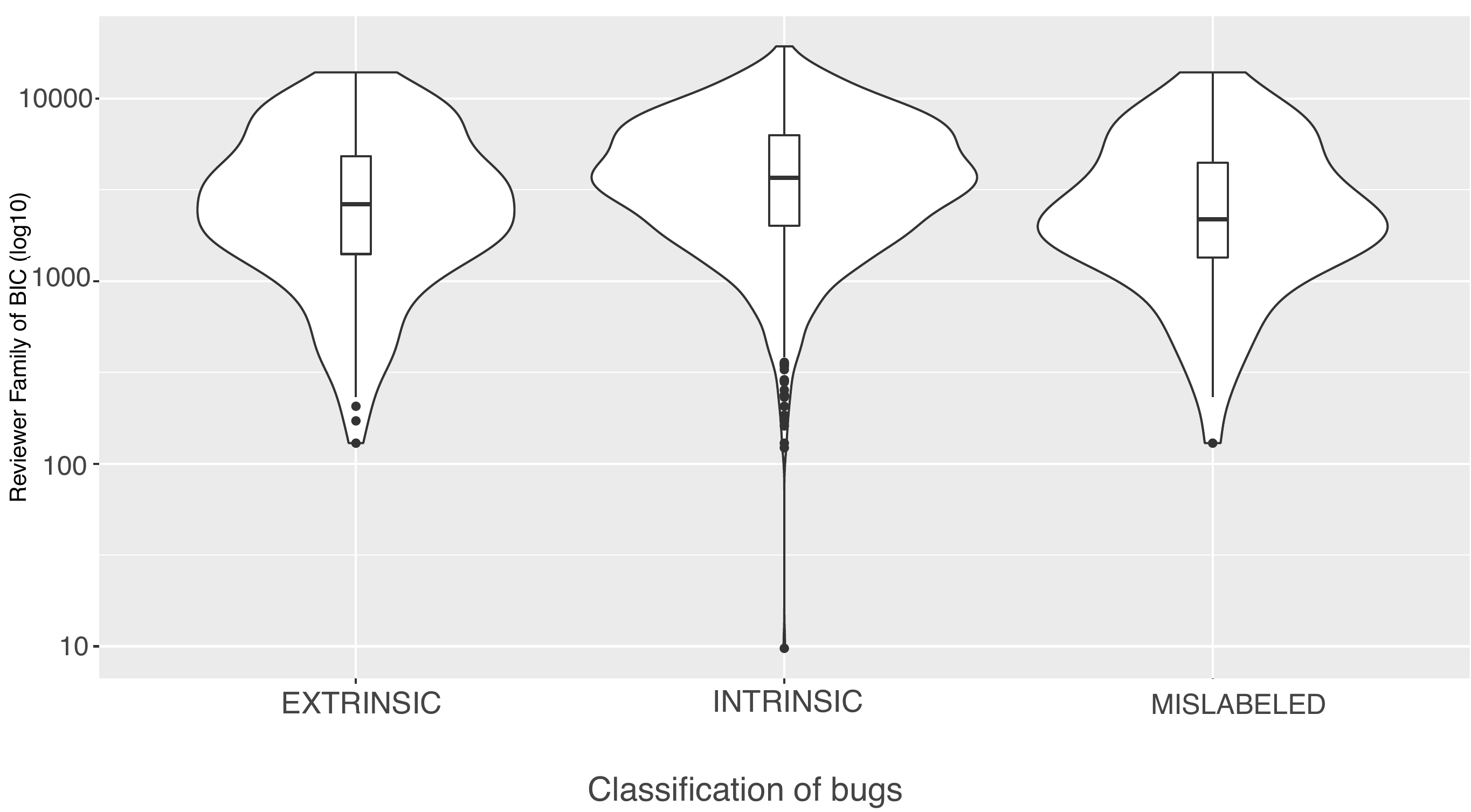}
		\caption[]%
		{{\small Reviewer}}
		\label{fig:reviewer}
	\end{subfigure}
	\hfill
	\begin{subfigure}[b]{0.33\textwidth}
		\centering 
		\includegraphics[width=\textwidth]{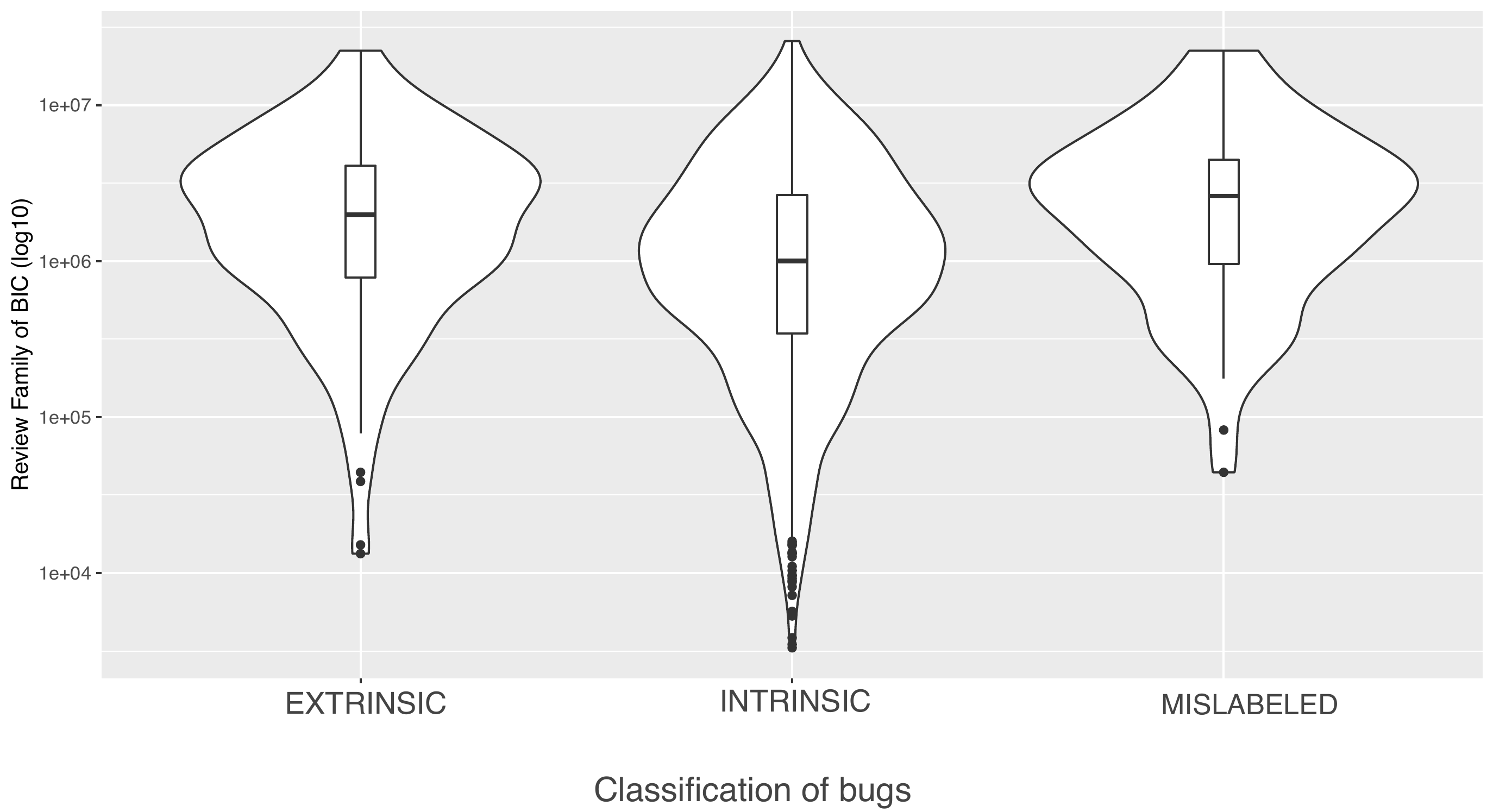}
		\caption[]%
		{{\small Review}}
		\label{fig:review}
	\end{subfigure}
	\caption[]
	{\small Distribution of intrinsic bugs, extrinsic bugs, mislabeled bugs and all bugs for the six families of code change properties. The families of code change properties are shown in Table~\ref{taxonomy}.} 
	\label{fig:distribution}
\end{figure*}

Figure~\ref{fig:distribution} shows violin plots with the distribution for each family of code change properties of the manually classified intrinsic, extrinsic, and mislabeled issues. The kernel plot indicates the distribution shape of the data. Wider sections represent a higher probability that members of the population will take on the given value; skinnier sections represent a lower probability.

Figure~\ref{fig:distribution} shows the distribution shape among the three groups per family of code change properties. Figure~\ref{fig:size} shows a bimodal distribution for extrinsic bugs.
The distribution shape of intrinsic and mislabeled bug is however uni-modal.
Besides, Figure~\ref{fig:diffusion} shows that the distribution frequency of intrinsic bugs are concentrated in lower values while the distribution frequency for extrinsic and mislabeled bugs is more uniform.

Figure~\ref{fig:distribution} offers evidence that (1) intrinsic and extrinsic bugs have different distributions and medians for all the six families; (2) intrinsic and mislabeled bugs also have a different distribution and medians; and (3) extrinsic and mislabeled bugs are more similar than intrinsic and mislabeled bugs in terms of distributions shape and median.

After computing the Kruskal‐Wallis test for the six families of code change properties, we obtained $p-values < 0.05$ in five of them. Thus, the \emph{Size} (1.9E.-014), \emph{Diffusion} (2.2E.-16), \emph{Reviewer} (1.6E.-06), \emph{Author} (2.6E.-05), and \emph{Review} (0.0005) families can be considered different for BICs linked to extrinsic, intrinsic and mislabeled bugs.

Table~\ref{wilcoxonBIC} shows which pairs of groups are different for the six families of BIC code change properties. This table offers evidence that the differences between intrinsic bugs and extrinsic bugs or mislabeled bugs are statistically significant for five out of six families. Furthermore, this table also points out that extrinsic bugs and mislabeled bugs are similar in four out of six families, i.e., \emph{Author}, \emph{Reviewer}, \emph{History}, and \emph{Review}. This finding illustrates that (1) intrinsic, extrinsic, and mislabeled bugs are not the same; and (2) extrinsic bugs and mislabeled bugs have code change properties that are very similar.

\begin{table}
	\renewcommand{\arraystretch}{1.4}
	\caption{p-values between Bug-Introducing Changes linked to different kind of bugs (Wilcoxon rank sum test).}
	\label{wilcoxonBIC}
	\centering
	\begin{smaller}
		\begin{tabular}{l@{ }|l@{ }|l@{ }|l@{ }|l@{ }|l@{ }|l@{ }}
			\toprule
			\textbf{} & \textbf{Size} & \textbf{Diffusion} & \textbf{Author} & \textbf{Reviewer}  & \textbf{History} & \textbf{Review} \\
			\hline
			Int.-Ext. & {\bf 2.4E.-09}&{\bf 1.4E.-13}&{\bf 0.001}& {\bf 0.0002} &0.9&{\bf 0.005}\\
			Int.-Mis. &  {\bf 4.8E.-08}&{\bf 3.1E.-02}&{\bf 0.002}&{\bf 0.002} &0.9&{\bf 0.012}\\
			Ext.-Mis. &  {\bf 2.7E.-01}&{\bf 4.8E.-14}&0.6& 0.5 &0.9&0.7\\
			\hline
			\bottomrule 
		\end{tabular}
	\end{smaller}
\end{table}

\textbf{Results}
    \textbf{b) Code Change Properties of BFCs for intrinsic, extrinsic, and mislabeled bugs}

After computing the Kruskal‐Wallis test for the six families code change properties of BFCs, we obtained $p-values < 0.05$ in three of them. Thus, the \emph{Size} ($4.8^-016$), \emph{Reviewer} ($0.0003$), and \emph{Author} ($0.002$) families can be considered different for BFCs linked to extrinsic, intrinsic and mislabeled bugs.

Table~\ref{wilcoxonBFC} shows which pairs of groups are different for the six families of BIC code change properties. This table offers evidence that (1) the differences between intrinsic and extrinsic bugs are statistically significant for \emph{Size} and \emph{Author}, (2) the differences between intrinsic and mislabeled bugs are statistically significant for \emph{Size}, \emph{Author}, \emph{Reviewer} and \emph{Review}; and (3) the differences between extrinsic bugs and mislabeled bugs are statistically significant for the \emph{Size} family.

\begin{table}
	\renewcommand{\arraystretch}{1.4}
	\caption{p-values between Bug-Fixing Changes linked to different kind of bugs (Wilcoxon rank sum test).}
	\label{wilcoxonBFC}
	\centering
	\begin{smaller}
		\begin{tabular}{l@{ }|l@{ }|l@{ }|l@{ }|l@{ }|l@{ }|l@{ }}
			\toprule
			\textbf{} & \textbf{Size} & \textbf{Diffusion} & \textbf{Author} & \textbf{Reviewer} & \textbf{History} & \textbf{Review} \\
			\hline
			Int.-Ext. & {\bf 0.02} &0.15&{\bf 0.043}&0.05&0.15&0.113\\
			Int.-Mis. & \textbf{4.9E.-16} &0.71&{\bf 0.005}& {\bf 0.0004} &0.17&{\bf 0.03}\\
			Ext.-Mis. & {\bf 0.005}&0.15&0.95& 0.61 &0.38&0.96\\
			\hline
			\bottomrule 
		\end{tabular}
	\end{smaller}
\end{table}

\vspace{0.2cm}
\fbox{\begin{minipage}{23em}
		\textbf{Answer to RQ6:} 
        Intrinsic and extrinsic bugs have different code change properties. When analyzing
mislabeled bugs as well, we have found that the \emph{nature} of extrinsic bugs is closer to them than to intrinsic bugs.
		These differences are statistically significant in five out of six families for BICs.
		For BFCs, half of code change families are statistically different.
\end{minipage}}

\section{Discussion and further research}
\label{sec:discussion}

In this section, we discuss the impact of our results first on JIT models and then on software engineering practice in general. We also discuss the implications of our results for researchers and practitioners.

\subsection{Impact on JIT}
\label{subsec:impact-jit}
Our results show that JIT models fed exclusively with intrinsic bugs obtain a more accurate representation of the real world; issues that are mislabeled bugs and bug reports that are due to extrinsic bugs should be removed.

The impact of this finding is significant, as over the past 15 years many studies
have used automatic techniques to collect bug datasets which are formed by bug reports, bug-fixing commits, and bug-introducing changes. These dataset are then used to train bug prediction models~\cite{zimmermann2007predicting,kim2008classifying,kim2007predicting,kamei2016studying,di2018developer,huang2018revisiting, pascarella2019fine}.

Hence, the results of hundreds of studies on bug prediction~\cite{hall2012systematic} may be not as accurate as they could, as
they have not discriminated between intrinsic and extrinsic bugs when training their models. 

On the other hand, our results support some of Mc\&K's results for JIT models.
When JIT models are trained without extrinsic bugs, we found that (1) they lose a large amount of predictive power one year after being trained; (2) when trained using periods that are closer to the testing period tend to outperform models that are trained using older periods; (3) long period JIT models do not always retain more predictive power for longer than short period JIT models; and (4) the \emph{Size} family is consistently the top contributor in our JIT models, and fluctuations in short period JIT models are more common than in long period JIT models.

As already mentioned, there is a debate in the research literature on the impact of mislabeled bugs. Some authors found that they introduce noise in the results of prediction models~\cite{kim2011dealing,herzig2013s}, while others contradict this finding~\cite{tantithamthavorn2015impact}. Our paper sheds more light on this topic, as it offers evidence that JIT models that used only intrinsic bugs obtained a more accurate representation of bugs, as intrinsic code change metrics fit better in JIT models. We believe that researchers should be aware of the noise that mislabeled bugs introduce in their dataset. Furthermore, we believe that a necessary criteria to assess the quality of the dataset is to select projects based on their policies to distinguish between bugs and non-bugs.


While mislabeled bugs have been widely studied in previous works~\cite{herzig2013s,rodriguez2016bugtracking,maalej2015bug}, extrinsic bugs have been recently discovered~\cite{rodriguez2019how} and we still do not understand them fully.
In our opinion, further research should be devoted to them.
It should be noted that in our case study the effect on 212 extrinsic bugs is similar to the one of 568 mislabeled bugs. Future research lines should continue studying how the characteristics of extrinsic bugs impact not just JIT bug prediction models, but also other bug prediction techniques.

\subsection{Impact on Software Engineering}
\label{subsec:impact-se}

We knew that not all the bugs are the same; they could be intrinsic or extrinsic depending of their origin~\cite{rodriguez2018if, rodriguez2019how}.
In this paper, we offer evidence that intrinsic and extrinsic have different code change properties.

We have also found more similarity between extrinsic bugs and mislabeled bugs in the patterns shown in Figure~\ref{fig:distribution}.
This result is also supported by Table~\ref{wilcoxonBIC} and Table~\ref{wilcoxonBFC} which indicate similar code changes properties between mislabeled bugs and extrinsic when analyzing BFCs and BICs linked to these bugs.
Furthermore, we have observed that some code change properties, i.e., \emph{Author}, \emph{History}, \emph{Review}, and \emph{Reviewer} of the BFC linked to extrinsic bugs and mislabeled bugs are similar. This finding might suggest that fixing a bug which does not have a BIC in the VCS can be compared to developing other kind of issues such as a mislabeled bug. In short, extrinsic bugs have similar characteristics than non-buggy changes. We find this evidence worth further research in order to understand the different \emph{natures} of bugs, and in particular extrinsic bugs.

We think our findings might have a broader impact than just improving bug prediction models. 

\textbf{Practices and processes} In the paper we have seen that, when removing extrinsic bugs, the explanatory power of the \emph{Size} family increases from 11\%-43\% to 20\%-49\%, but the explanatory power of the \emph{Review} family decreases considerably from 2\%-59\% to 6\%-21\% (see RQ3).
This points out that review practices may affect extrinsic and intrinsic bugs in a different manner, and thus should be addressed differently. 
In this regard, it would be interesting to see if there are practices that minimize the number (or at least the effect) of extrinsic bugs. We imagine as well that some software architectures could be more robust than others.

\textbf{Education} We believe that there is currently a strong bias towards training future software engineers exclusively on intrinsic bugs when identifying the origin of bugs as previous studies do not consider the extrinsic nature of bugs~\cite{rodriguez2018if,rodriguez2019how}. Our findings suggest that we should educate students in the fact that software bugs do not always have their origin in a change in the VCS. If tools and practices to support bug fixing of extrinsic bugs appear, we should incorporate them to the curriculum.

\subsection{Implications}

Besides the impact that our results have on JIT models and Software Engineering, we discuss the implications for developers, researchers, and practitioners.

\textbf{Data Awareness:}  If researchers include all bugs in their datasets,   they are using a dataset which has not been conveniently prepared, and the results could differ from reality. Thus, if developers are aware of the type of bug they are fixing and start labeling them accordingly in bug tracking systems or commit messages, researchers could obtain better datasets for bug prediction models and foster research on this issue. We hypothesize that software projects will benefit from this as well in the long run. 

In the past, we had a similar situation when developers started to indicate the ITS bug id in the BFC; this helped considerably in the improvement of the SZZ algorithm~\cite{bachmann2010missing}. ITSs also offer the possibility to categorize issues as mislabeled bugs. At this point, we do not if our results may lead to a drastic changes for developers because with one case study we are not able generalize. But, in the case of OpenStack the models without extrinsic bugs perform usually slightly better, sometimes much better.

Furthermore, researchers should be aware of their data and put more attention in the data collection process. They must ensure that when gathering data the ITSs selected for the study distinguish between bug reports and other kind of issues. Therefore, data validation is recommended~\cite{herzig2013s}.

\textbf{Tools:} The curation of bugs is a labor-intensive task that requires expert knowledge of the software system, which makes it a very costly process. Thus, the development of tools that help in the classification of bugs might be useful for researchers. In the same manner as tools have been developed that help to lower mislabeling~\cite{bachmann2010missing,wu2011relink,wen2016locus}, new tools could automatically detect intrinsic and extrinsic bugs. These tools can help practitioners and researchers to ensure the maintainability of software systems, nonetheless the quality of the datasets used to train bug prediction models.
For example, a new search could be how to use natural language processing techniques in combination with deep learning techniques to classify bugs as extrinsic or intrinsic based on the textual information from the bug reports. Also, another research line could study different techniques to automate as much as possible the theoretical model proposed by Rodr\'iguez-P\'erez \emph{et al.}~\cite{rodriguez2019how} to identify extrinsic and intrinsic bugs. 
We envision that these tools might be of benefit in other fields of software engineering such as testing/verification, software analytics and software maintenance and evolution.

\textbf{Research:} The different nature of extrinsic bugs compared to intrinsic ones demands as well further research; based on our previous work~\cite{rodriguez2018if,rodriguez2019how}, we conjecture that previous studies have focused much on the latter, but there is a lack of understanding, research and tools on the former. We call for more investigations on the topic. We need to know more about extrinsic bugs. We know very little about them. Are there different types of extrinsic bugs? Are they more costly than intrinsic bugs? Are they re-opened more often? Can we write software that is less prone to contain extrinsic bugs? Our aim with this paper has been not to focus only on the impact on extrinsic bugs on JIT bug prediction, but to draw attention to the fact that there is a new field of research in knowing more about extrinsic bugs.


\section{Threats to Validity}
\label{sec:threats}

The validity of this study is described in terms of the three main threats to validity that affect empirical software engineering research: construct, internal, and external~\cite{wohlin2012experimentation}.

\textbf{Construct Validity}
Since we are using the replication package provided in Mc\&K's paper~\cite{mcintosh2018fix}, this study suffers from the same construct validity threats reported in Mc\&K's study.
We have attempted to mitigate some of these threats.
For example, they used the SZZ algorithm to identify BICs without further refinement.
SZZ is widely used algorithm in bug prediction research~\cite{kamei2013large,kim2008classifying,kim2007predicting}, but it is well-known that it suffers from several limitations~\cite{da2017framework,rodriguez2018reproducibility}.
In this work, we manually identify those issues that are not related to a bug and then discriminate between extrinsic and intrinsic bugs.
Only for the latter the use of SZZ makes sense. 
The classification of 705 issues by only a single rater can be a threat to the validity. However, we tried to minimize the impact of this threat by training the two raters until they achieved a near perfect agreement before they starting classifying the 705 issues. This training include the analysis of 470 issues (25\%).

\textbf{Internal Validity}
Although we have experience in OpenStack from investigating it for several years, we have no advanced development expertise in this system.
This fact may have influenced the manual classification of bugs into the different types.
To mitigate this threat, we discussed the unclear cases, and when no agreement was reached, we treated these cases as Mc\&K's paper did (i.e., we considered that these bug reports were ``true" bug reports and not other kind of issues).

\textbf{External Validity}
A notable difference between Mc\&K's study and ours is that they had two case studies (OpenStack and Qt), while we have only one (OpenStack).
The rationale for this is that our study is very labor-intensive; while Mc\&K apply directly SZZ to the dataset of 1,880 issues, we have curated them manually.
The curation procedure requires to understand the bug in its very detail, which is a non-trivial task.
In total, raters have devoted over 250 hours carrying out the task of classifying these 1,880 issues.
The study of just one case study prevents us to generalize our findings to other systems.
However, our goal was not claim that our results would stand to all systems, but rather to show the exception, we have found that extrinsic bugs can have a significant impact on bug prediction models, at least in one project. 
We think that our research is successful in this regard, as we demonstrate that intrinsic and extrinsic bugs show different characteristics.
In the particular case of JIT models, we offer sufficient evidence that researchers and practitioners should be aware of extrinsic bugs (in addition to mislabeled bugs). Case studies contribute to increase knowledge and gain a deep understanding of particular phenomenon~\cite{runeson2012case}. Also, some theorist argument that case studies help to draw attention to things that need change~\cite{easterbrook2008selecting}.

\section{Conclusions}
\label{sec:conclusions}




Previous studies on Just In Time (JIT) bug prediction have not only assumed that future BICs are similar to past ones, but also that all bugs from the project can be linked to explicit BICs. As the research literature has shown~\cite{rodriguez2018if}, this does not always happen. Often it is not possible to find a BIC for a bug fix. Those bugs are referred to as extrinsic bugs, and are mainly caused by external factors to the project, such as changes to APIs or changes in the requirements.

Through a case study of the OpenStack system, we have investigated whether extrinsic bugs have an impact on JIT models. Our results indicate the negative role that extrinsic bugs have on the performance of JIT approaches.
When removing extrinsic bugs from the trained data used in OpenStack, JIT 
models obtain a more accurate representation of the real world as indicated by their different (often higher) AUC values in their performance. These models capture change properties better. Therefore, JIT models that are fitted only with intrinsic bugs obtain more stable AUC scores and lose less predictive power.



Our findings also support in part McIntosh and Kamei's results~\cite{mcintosh2018fix}. We found that after removing extrinsic bugs, the values of the importance score of the six source code change families fluctuate as the system evolves and that these fluctuations can lead to underestimate or overestimate the future impact of those families. 

Researchers and practitioners should be aware of the data that feed JIT bug prediction models. They should perform data validation to ensure that only intrinsic bugs are considered when training their models.
Although with the current state of the art data validation might be tedious and very labor-intensive to achieve, at least researchers should be aware that considering extrinsic bugs during the training of the models might impact bug prediction results.

All in all, we show evidence that extrinsic bugs are of different \emph{nature} than intrinsic bugs.
Actually, they are more similar to issues that are not bugs than to intrinsic bugs.
We think that this finding is not only relevant for JIT bug prediction models, but that it may impact many other areas of software engineering practice and research, and would like to call for further research on extrinsic bugs.

A future line of research will be the semi-automation of the process to identify extrinsic bugs. Our experience shows that this will not be an easy process because researchers have to understand at least the bug description (in natural language) and the change (code). We envision that a semi-automated process will require the combination of different techniques and tools. For example, to understand the bug description researchers can implement natural language processing; and to understand the source code they can use tools that help researchers to backtrack the evolution of source code lines from their introduction in the file until their modification in the bug fixing commit. More details can be found in our replication package.

\textbf{Replication package}: We have set up a replication package\footnote{\url{http://gemarodri.github.io/2019-Study-of-Extrinsic-Bugs/}} including data sources, intermediate data, and scripts.

\ifCLASSOPTIONcompsoc
  \section*{Acknowledgments}
\else
  \section*{Acknowledgment}
\fi
We acknowledge the support of the Government of Spain through project ``BugBirth'' (RTI2018-101963-B-100). Many thanks to JJ~Merchante, who devoted tons of hours to classify bugs. We would like to thank as well McIntosh and Kamei for their invaluable support, and J.M. Gonzalez-Barahona, A. Serebrenik, A. Zaidman, and D.M. German for their invaluable feedback and discussions.

\bibliography{bibliography.bib}
\bibliographystyle{ieeetr}

\end{document}